\documentstyle[12pt,epsf]{article}

\def\etal{{\it et al.}}

\def\abs#1{\left| #1\right|}
\def\sgn{\mathop{\rm sgn}}
\def\etmiss{\slashchar{E}_T}
\def\fb{{\rm fb}}

\def\tG{{\tilde G}}
\def\ns{{\rm ns}}
\def\tell{{\tilde\ell}}
\def\ttau{{\tilde\tau}}

\def\Meff{M_{\rm eff}}

\def\lsp{{\tilde\chi_1^0}}

\def\GeV{{\rm GeV}}

\def\tchi{\tilde\chi}
\def\tg{\tilde g}
\def\tq{\tilde q}
\def\Cgrav{C_{\rm grav}}
\let\badcite=\cite
\def\cite{~\badcite}

\def\Frac#1#2{{\displaystyle#1\over\displaystyle#2}}

\def\slashchar#1{\setbox0=\hbox{$#1$}           
   \dimen0=\wd0                                 
   \setbox1=\hbox{/} \dimen1=\wd1               
   \ifdim\dimen0>\dimen1                        
      \rlap{\hbox to \dimen0{\hfil/\hfil}}      
      #1                                        
   \else                                        
      \rlap{\hbox to \dimen1{\hfil$#1$\hfil}}   
      /                                         
   \fi}                                         %

\catcode`@=11
\newdimen\vbigd@men                             

\def\vbigl{\mathopen\vbig}

\def\vbigr{\mathclose\vbig}

\def\vbig#1#2{{\vbigd@men=#2\divide\vbigd@men by 2%
   \hbox{$\left#1\vbox to \vbigd@men{}\right.\n@space$}}}
\catcode`@=12

\catcode`@=11
\def\citenum#1{\csname b@#1\endcsname}
\catcode`@=12

\def\dofig#1#2{\centerline{\epsfxsize=#1\epsfbox{#2}}}
\begin{document}
\begin{titlepage}
\rightline{LBNL-42401}
\rightline{BNL-HET-98/37}
\rightline{ATL-PHYS-98-134}

\bigskip\bigskip

\begin{center}{\Large\bf Measurements in Gauge Mediated SUSY Breaking
  Models at LHC\footnotemark}
\end{center}
\footnotetext{This work was supported in part by the Director, Office
of Energy Research, Office of High Energy and Nuclear Physics,
Division of High Energy Physics of the U.S. Department of Energy under
Contracts DE-AC03-76SF00098 and DE-AC02-98CH10886.}

\bigskip
\centerline{\bf I. Hinchliffe$^a$ and F.E. Paige$^b$
}
\centerline{$^a${\it Lawrence Berkeley National Laboratory, Berkeley, CA}}
\centerline{$^b${\it Brookhaven National Laboratory, Upton, NY}}
\bigskip

\begin{abstract}
Characteristic examples are presented of scenarios of particle
production and decay in supersymmetry models in which the
supersymmetry breaking is transmitted to the observable world via
gauge interactions. The cases are chosen to illustrate the main
classes of LHC phenomenology that can arise in these models. A new 
technique is illustrated that allows the full reconstruction 
of supersymmetry events despite the presence of two unobserved particles.
This technique enables superparticle masses to be measured directly rather than
being inferred from kinematic distributions.
It is demonstrated that the LHC is capable of making sufficient
measurements so as to  severely over-constrain the model and  determine the
parameters with great precision.
\end{abstract}
\end{titlepage}

\tableofcontents

\newpage
\section{Introduction}
\label{sec:intro}

If supersymmetry (SUSY) exists at the electroweak scale, then
gluinos and squarks will be copiously produced in pairs at the LHC and
will decay via cascades involving other SUSY particles. The
characteristics of these decays and hence of the signals that will
be observed and the measurements that will be made depend upon the
actual SUSY model and in particular on the pattern of supersymmetry
breaking.
Previous, detailed studies of signals for SUSY at the
LHC\cite{hinch,previous,fabiola} have used the SUGRA model 
\cite{SUGRA,SUGRArev}, 
in which the supersymmetry breaking is transmitted to the sector of the 
theory containing the Standard Model particles and their superpartners 
via gravitational interactions. The minimal version of this model has
just four parameters plus a sign.
The lightest supersymmetric particle ($\lsp$) has a mass of order
100~GeV, is stable, is produced in the decay of every other supersymmetric 
particle and  is  neutral and therefore escapes the detector. The
strong production cross sections and the characteristic
signals of events with multiple jets plus missing energy $\etmiss$ or with
like-sign dileptons $\ell^\pm\ell^\pm$ plus
$\etmiss$\cite{BCPT} enable SUSY to be extracted trivially
from Standard Model backgrounds. Characteristic signals were
identified that can be exploited to determine, with great precision,
the fundamental
parameters of these model over the whole of its parameter space.
Variants of this model where 
R-Parity is broken\cite{hall} and the  $\lsp$ is
unstable have also been discussed\cite{rparity}. 

There also exists a class of gauge mediated SUSY breaking (GMSB)
models\cite{GMSB, GMSBrev} 
in which the supersymmetry breaking is 
mediated by gauge interactions. The model assumes that supersymmetry
is broken with a scale $\sqrt{F}$ in a sector of the theory which
contains heavy non-Standard-Model particles. This sector then couples
to a set of particles with Standard Model interactions,
called messengers, which have a
mass of order $M$. These messengers are taken to be complete
representations of SU(5) so as to preserve the coupling constant
unification of the Minimal Supersymmetric Standard Model
(MSSM). The mass splitting between the superpartners in the messenger
multiplets is controlled by $\sqrt{F}$. One (two) loop graphs involving these 
messenger fields, then give mass to superpartners of the gauge bosons
(quarks and leptons) of the Standard Model. This model is 
preferred by 
some because the superpartners of the Standard Model particles get
their masses via gauge interactions, so there are no flavor changing
neutral currents, which can be problematic in the SUGRA models.

The characteristic spectra of
superparticles are different from those in the SUGRA model;
in particular,  the lightest supersymmetric particle
is now the gravitino ($\tilde{G}$). This particle has feeble
interactions and can be produced with significant rate only in the
decays of particles which have no other decay channels. In the SUGRA
models, $\tilde{G}$ has a mass of order 1 TeV and is phenomenologically 
irrelevant (except possibly for cosmology). 
In GMSB models, the next lightest supersymmetric particle, 
which we will refer to as the NLSP, decays into $\tilde{G}$. The
lifetime of the NLSP  is very model dependent:  $ O(1 \mu{\rm m})<
c\tau < O({\rm many\ km})$. As it is not stable, it can either be
charged or neutral. 

If the NLSP is neutral, it is the lightest combination ($\tilde\chi_1^0$) 
of gauginos and Higgsinos, and it behaves, except for its decay,
in the same manner as  the $\lsp$ in the SUGRA model. (The rather unlikely
possibility that it is a gluino\cite{gluino-lsp} 
is not explored in this paper.) If its 
lifetime is very long so that none of the produced $\lsp$ decay
within the detector, the phenomenology is very similar to that in the
SUGRA models.

If the NLSP  is charged, it is most likely to be 
 the partner of a right handed
lepton. Two cases are distinguished. If $\tan \beta$, the ratio of the 
vacuum expectation values of the two Higgs fields, is small, then
$\tilde{e}_R$, $\tilde{\mu}_R$ and $\tilde{\tau}_R$ are degenerate. If
$\tan\beta$ is large,  $\tilde{\tau}_R$ is the lightest slepton and the 
others can decay into it with short lifetimes. If the lifetime of the NLSP
is very long,  each SUSY event contains two apparently stable 
charged particles \cite{feng}. If it is short, each event contains two charged
leptons from each of the decays.

The simulation in this paper is based on the implementation of the
minimal GMSB model in ISAJET\cite{ISAJET}.
The model
is characterized by $\Lambda = F/M_m$, the
SUSY breaking mass scale; $M_m$, the messenger mass; $N_5$, the number
of equivalent $5+\bar5$ messenger fields; $\tan\beta$, the usual ratio
of vacuum expectation values of the Higgs fields that couple to the
charge 2/3 and 1/3 quarks; $\sgn\mu=\pm1$, the sign of the $\mu$
term, the value $\abs{\mu}$ being determined  
by the $Z$ mass from usual radiative electroweak symmetry
breaking; and $\Cgrav\ge1$, the scale factor for the gravitino mass
which determines the NLSP lifetime ($\tau_{\rm NLSP}\sim \Cgrav^2$).
At the scale $M_m$, for example, the masses of the gluino, 
squark and slepton are given by
\begin{eqnarray*}
m_{\tilde{g}}&=& \Frac{\alpha_s}{4\pi}\Lambda N_5\\
m_{\tilde{e}_L}^2&=&\Frac{3\alpha_2^2}{32\pi^2}\Lambda^2N_5
+\Frac{3\alpha_1^2}{160\pi^2}\Lambda^2N_5\\
m_{\tilde{e}_R}^2&=&\Frac{3\alpha_1^2}{40\pi^2}\Lambda^2N_5\\
m_{\tilde{u}_L}^2&=&\Frac{\alpha_s^2}{6\pi^2}\Lambda^2N_5+
\Frac{3\alpha_2^2}{32\pi^2}\Lambda^2N_5+
\Frac{\alpha_1^2}{480\pi^2}\Lambda^2N_5\\
m_{\tilde{u}_R}^2&=&\Frac{\alpha_s^2}{6\pi^2}\Lambda^2N_5
+\Frac{\alpha_1^2}{30\pi^2}\Lambda^2N_5
\end{eqnarray*}
Here $\alpha_s$, $\alpha_2$ and $\alpha_1$ are the coupling constants of
$SU(3)$, $SU(2)$, and $U(1)$ respectively. These masses are then
evolved from 
the scale $M_m$ to the weak scale, inducing a logarithmic dependence on $M_m$.
As in the case of the SUGRA models, this evolution leads to the
spontaneous breaking of electro-weak symmetry as the large top quark
Yukawa coupling of the top (and possibly bottom) quark induces a
negative mass-squared of the Higgs field(s).
From these equations it can be seen that as $N_5$
 is increased the slepton masses increase more slowly than the gaugino
 masses.  Hence only 
for small values of $N_5$ will the NLSP be a $\tchi_1^0$, at larger
values it  
is a (right) slepton.
Messenger fields in other representations of SU(5) can be included by
modifying the value of $N_5$. If there are several messengers with
 different masses their effects can be approximated by changing $N_5$. 
Therefore, we will consider $N_5$ to be a
continuous variable when we estimate how well it can be measured at
LHC.

This paper presents a series of case studies for this model which
illustrate its characteristic features. We use the  four sets of
parameters shown in Table 1. These cases are paired and differ only in
the value of  $\Cgrav$ and hence in the lifetime of the NLSP. The
masses of the superpartners of the Standard Model fields are given in
Table 2. In cases G1a and G1b, the NLSP is $\chi_1^0$; its lifetime is 
quite short, $c\tau=1.2\,{\rm mm}$, in the former case and long,
$c\tau=1$~km, in the
latter. In cases G2a and G2b the NLSP is a
stau. In the latter case it has a very long lifetime, $c\tau \approx
1.25\,{\rm km}$, and exits a detector without decaying.  
The decay $\tell \to \ttau_1
\tau \ell$ is not kinematically allowed, so the $\tilde e_R$ and
$\tilde\mu_R$ are also long lived. In the case of point G2a, the
$\tilde{\tau}_1$, $\tilde e_R$ and $\tilde\mu_R$ are short lived with
$c\tau=52$ $\mu$m. 
The production cross section for supersymmetric particles is quite large; 
7.6 pb and 22 pb at lowest order in QCD for cases G1 and G2 respectively.
Note that the larger value of $N_5$ in the case of G2 results in considerably 
smaller squark masses and hence the larger cross section; the gluino
masses are very similar. 

\begin{table}[t]
\caption{Parameters for the four case studies in this paper.\label{params}} 
\begin{center}
\begin{tabular}{ccccccc}
\hline\hline
Point & $\Lambda$ & $M_m$ & $N_5$ & $\tan\beta$ & $\sgn{\mu}$&$\Cgrav\ge1$  \\
      & (TeV) & (TeV)   &    &             &             \\
\hline

G1a & 90 & 500 &   1 & \phantom{0}5.0 & $+$&$1.0$ \\
G1b & 90 & 500 &   1 & \phantom{0}5.0 & $+$&$10^3$ \\
G2a & 30 & 250 &   3 & \phantom{0}5.0 & $+$&$1.0$ \\
G2b & 30 & 250 &   3 & \phantom{0}5.0 & $+$&$5\times 10^3$ \\
\hline\hline
\end{tabular}
\end{center}
\end{table}

\begin{table}[t]
\caption{Masses of the superpartners, in GeV,  for the cases to be studied.
  Note that the first
and second generation squarks and sleptons are degenerate and so are 
not listed separately. \label{mass-table}} 
\begin{center}
\begin{tabular}{ccccccc} \hline \hline 
Sparticle & G1 & G2 & \qquad &Sparticle & G1 & G2\\ \hline
$\widetilde g$  & 747& 713 &&&& \\
 $\widetilde \chi_1^\pm$  &  223 &  201 &&
$\widetilde \chi_2^\pm$        &  469&  346 \\
 $\widetilde \chi_1^0$          &  119 &  116 & &
$\widetilde \chi_2^0$          &  224 &  204  \\
 $\widetilde \chi_3^0$          &  451 &  305  &&
$\widetilde \chi_4^0$          &  470 &  348  \\
$\widetilde u_L$               &  986 &  672  &&
$\widetilde u_R$               &  942 &  649  \\
$\widetilde d_L$               &  989 &  676  &&
$\widetilde d_R$               &  939 &  648  \\
$\widetilde t_1$               &  846 &  584  &&
$\widetilde t_2$               &  962 &  684  \\
$\widetilde b_1$               &  935 &  643  &&
$\widetilde b_2$               &  945 &  652  \\
$\widetilde e_L$               &  326 &  204  &&
$\widetilde e_R$               &  164 &  103  \\
$\widetilde \nu_e$             &  317 &  189  &&
$\widetilde \tau_2$            &  326 &  204  \\
$\widetilde \tau_1$            &  163 &  102  &&
$\widetilde \nu_\tau$          &  316 &  189  \\
$h^0$                          &  110 &  107  &&
$H^0$                          & 557 &  360 \\
$A^0$                          & 555 &  358 &&
$H^\pm$                        & 562 &  367 \\
\hline \hline
\end{tabular}
\end{center}
\end{table}

        All the analyses presented here are based on
ISAJET~7.37\cite{ISAJET} and a simple detector simulation. At least 50K
events were generated for each signal point. The Standard Model
background samples contained 250K events for each of $t \bar t$, $WZ$
with $W \to e\nu,\mu\nu,\tau\nu$, and $Zj$ with $Z \to
\nu\bar\nu,\tau\tau$, and 5000K QCD jets (including $g$, $u$, $d$,
$s$, $c$, and $b$) divided among five bins covering $50 < p_T <
2400\,\GeV$. Fluctuations on the histograms reflect the generated
statistics.  On many of the plots that we show, very few Standard
Model background events survive the cuts and the corresponding
fluctuations are large, but in all cases we can be confident that the
signal is much larger than the residual background.  The cuts that we
choose have not been optimized, but rather have been chosen to get
background free samples.

        The detector response is parameterized by Gaussian resolutions
characteristic of the ATLAS\cite{ATLAS} detector without any tails.
All energy and momenta are measured in GeV.  In the central region of
rapidity we take separate resolutions for the electromagnetic (EMCAL)
and hadronic (HCAL) calorimeters, while the forward region uses a
common resolution:
\begin{eqnarray*}
{\rm EMCAL} &\quad& 10\%/\sqrt{E} \oplus 1\%, |\eta|\, <3 \,\nonumber\\
{\rm HCAL}  &\quad& 50\%/\sqrt{E} \oplus 3\%, |\eta|\, < 3\,\nonumber\\
{\rm FCAL}  &\quad& 100\%/\sqrt{E} \oplus 7\%, |\eta|\, > 3\,\nonumber
\end{eqnarray*} 
A uniform segmentation $\Delta\eta = \Delta\phi = 0.1$ is used with no
transverse shower spreading; this is particularly unrealistic for the
forward calorimeter. Both ATLAS\cite{ATLAS} and CMS\cite{CMS} have
finer segmentation over most of the rapidity range.  An oversimplified
parameterization of the muon momentum resolution of the ATLAS detector
including a both the inner tracker and the muon system measurements is
used, {\it viz}
$$
\delta p_T/p_T = \sqrt{0.016^2+(0.0011p_T)^2}
$$
In the case of electrons we take a momentum resolution obtained by
combining the electromagnetic calorimeter resolution above with a
tracking resolution of the form
$$
\delta p_T/p_T= \left(1+{0.4\over(3-\abs{\eta})^3}\right)
\sqrt{(0.0004p_T)^2+0.0001}
$$
This provides a slight improvement over the calorimeter alone.
Missing transverse energy is calculated by taking the magnitude of the
vector sum of the transverse energy deposited in in the calorimeter
cells. 

        Jets are found using GETJET\cite{ISAJET} with a simple
fixed-cone algorithm. The jet multiplicity in SUSY events is rather
large, so we will use a cone size of 
$$
R = \sqrt{(\Delta\eta)^2+(\Delta\phi)^2} = 0.4
$$ 
unless otherwise stated. Jets are required to have at least $p_T >
25$~GeV; more stringent cuts are often used.  All leptons are required
to be isolated and have some minimum $p_T$ and $\abs{\eta}< 2.5$.  The
isolation requirement is that no more than 10~GeV of additional $E_T$
be present in a cone of radius $R = 0.2$ around the lepton to reject
leptons from $b$-jets and $c$-jets. In addition to these kinematic
cuts a lepton ($e$ or $\mu$) efficiency of 90\% and a $b$-tagging
efficiency of 60\% is assumed\cite{ATLAS}. Where relevant, we include
the possibility that jets could appear as photons in the detector due
to fragmentation where most of the jet energy is taken up by
$\pi^0$'s.  Jets are picked randomly with a probability of $10^{-3}$.
They are then called photons and removed from the list of jets. This
is a conservative assumption, ATLAS is expected to have a better rejection

        Results are presented for an integrated luminosity of
$10\,\fb^{-1}$, corresponding to one year of running at $10^{33}\,{\rm
cm^{-2}s^{-1}}$ so pile up has not been included. We will occasionally
comment on the cases where the full design luminosity of the LHC, {\it
i.e.}\ $10^{34}\,{\rm cm^{-2}s^{-1}}$, will be needed to complete the
studies. For many of the histograms shown, a single event can give
rise to more than one entry due to different possible combinations.
When this occurs, all combinations are included.

The next sections of this paper contain detailed examples of
analyses that could be carried out in  the selected cases. In particular, we
illustrate a technique where the momenta of each of the $\tilde{G}$'s can be
reconstructed even though only the sum of their transverse momenta is
measured directly.
Using these momenta we are then able to reconstruct the 
squark and gluino decays. 
We devote considerable space to this technique as it is new
and enables the masses of superparticles to be directly measured rather 
than inferred from kinematic distributions. 
This technique should be applicable to other cases. 
We then 
estimate how well the LHC could determine the parameters of the model
and comment on  possible ambiguites in interpreting the signal. In those cases
where the signal is characteristally differant from the SUGRA cases, 
the expected precision on the model parameters is even better than that
in the SUGRA cases.  Finally we comment upon how our results can be used
to estimate how well the LHC would be able to study other parameter
sets of the GMSB model. 

\section{Point G1a}
\label{sec:meff}

In  case G1a, the NLSP is $\chi_1^0$ and decays  to
$\tG \gamma$ with $c\tau=1.2\,{\rm mm}$.
The total SUSY cross section is 7.6 pb. 
SUSY events are
characterized by two hard isolated photons plus the usual jets,
leptons, and missing transverse energy $\etmiss$. The presence of two
photons in almost every event renders the  Standard Model
backgrounds negligible. The first evidence
for new physics in this case will arise from a huge excess
 of events with two photons and missing $E_T$ over 
that expected from the Standard Model.

In a small fraction ($2.0\%$) of the
events, the NLSP will undergo a Dalitz decay to $e^+e^-\tG$. 
The electron and positron can be used to accurately determine
the decay vertex and a precise measurement of the lifetime made.
The systematic limit on the precision from the resolution of the
vertex system of  ATLAS  is at the 10$\mu$m  level. 
The precision will therefore be
limited by statistics of the thousand or so observed Dalitz decays.
This measurement is important as it provides the only constraint on
$\Cgrav$ and hence on the true scale of SUSY breaking.

The effective mass is defined to be the scalar sum of the missing
transverse energy $\etmiss$ and the $p_T$'s of the four hardest jets,
which are required to contain more than one charged particle with
$p_T>1\,\GeV$:
$$
\Meff \equiv \etmiss + p_{T,1} + p_{T,2} + p_{T,3} + p_{T,4}\,.
$$
This is a good measure of the hardness of the event. Events are
selected to have 
\begin{itemize}
\item   $\Meff > 400\,\GeV$;
\item   $\etmiss > 0.1\Meff$.
\end{itemize}
Electrons and photons are required to have $p_T>20\,\GeV$; muons are
required to have $p_T>5\,\GeV$. At least two photons and two leptons
are also required for all the analyses in this section.

\subsection{Lepton/Photon Distributions}
\label{sec:g1alep}

\begin{figure}[t]
\dofig{3.5in}{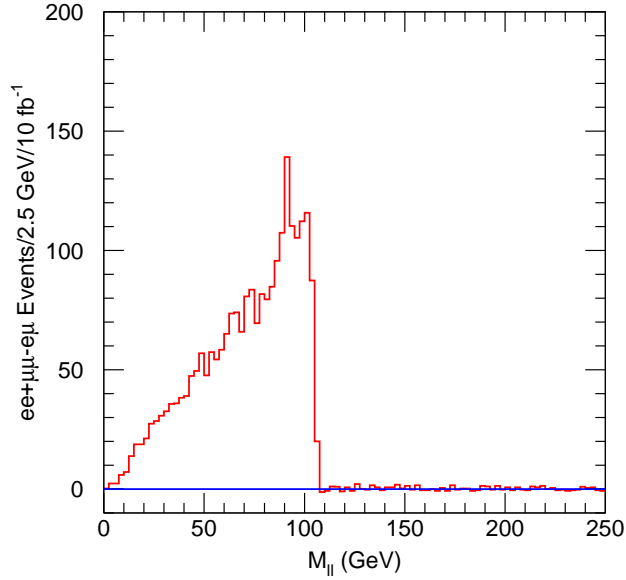}
\vskip-14pt
\caption{$M_{\ell\ell}$ distribution for $e^+e^- + \mu^+\mu^- -
e^\pm\mu^\mp$ events with two photons and two isolated leptons arising 
at Point G1a. The
Standard Model background is negligible.\label{mll}}
\end{figure}

We begin a detailed study by first selecting  events that have 
at least two leptons (either electrons or muons)
and two photons satisfying the cuts described above. In order to
cleanly select events arising from the decay chain 
$$
\tchi_2^0 \to \tell^\pm\ell^\mp \to \lsp\ell^\pm\ell^\mp \to
\tG\gamma\ell^\pm\ell^\mp\,,
$$ 
we observe that these leptons are correlated in flavor and hence we
take events where a pair of the leptons have opposite charge and form the
flavor subtracted combination 
 $e^+e^- + \mu^+\mu^- - e^\pm\mu^\mp$. The distribution in the
 $\ell^+\ell^-$ mass distribution
 is shown in Figure~\ref{mll}. 
This distribution has a sharp  edge at
$$
M_{\tchi_2^0} \sqrt{1-\left(M_{\tell_R} \over M_{\tchi_2^0}\right)^2}
\sqrt{1-\left(M_{\lsp} \over M_{\tell_R}\right)^2} = 105.1\,\GeV.
$$
which arises from the decay chain above. If  we do not form
 the flavor combination, the 
edge is still clearly visible but there is much more combinatorial 
background. The position of this edge can 
be used to determine this combination of masses with great precision
given the huge statistical sample. The ultimate limit will arise from
systematics in the measurement of dilepton mass distribution, which
given the large sample of $Z\to \ell^+\ell^-$ decays that will be
available for calibration, we expect to have an uncertainty of order
 0.1\% or 100 MeV.

\begin{figure}[t]
\dofig{3.5in}{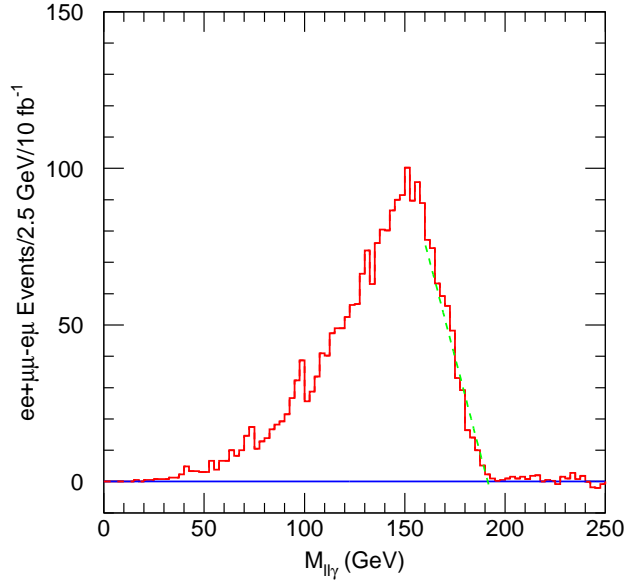}
\vskip-14pt
\caption{$M_{\ell\ell\gamma}$ distribution for $e^+e^- + \mu^+\mu^- -
e^\pm\mu^\mp$ events with two photons at Point~G1a. 
A linear fit from 160 to 190 GeV used to determine the position of the 
4-body endpoint is also shown as a dashed line.\label{mllg}}
\end{figure}

From the same sample of events with two leptons and two photons, the
smaller of the two $\ell^+\ell^-\gamma$ masses is
formed and the  resulting mass distribution is shown in Figure \ref{mllg}.
The endpoint is at 
$$
\sqrt{M_{\tchi_2^0}^2 - M_{\chi_1^0}^2}=189.7\,\GeV\,,
$$
the kinematic limit for $\tchi_2^0 \to \tG \ell^+\ell^-\gamma$.
Instead of having a sharp edge like Figure~\ref{mll}, this
distribution vanishes linearly because of four-body phase space.  The
figure shows as a dotted line a linear fit to the end of the spectrum.
From this fit the precise end point can be determined. The smaller
statistical sample and the need for a fit imply that the resulting
uncertainty on the value of the end point is larger. We estimate the
precision from Figure~\ref{mllg} to be $\pm 500$~MeV; the full LHC
luminosity of 100~fb$^{-1}$ should enable this uncertainty to be
reduced to $\pm 200$~MeV.

\begin{figure}[t]
\dofig{3.5in}{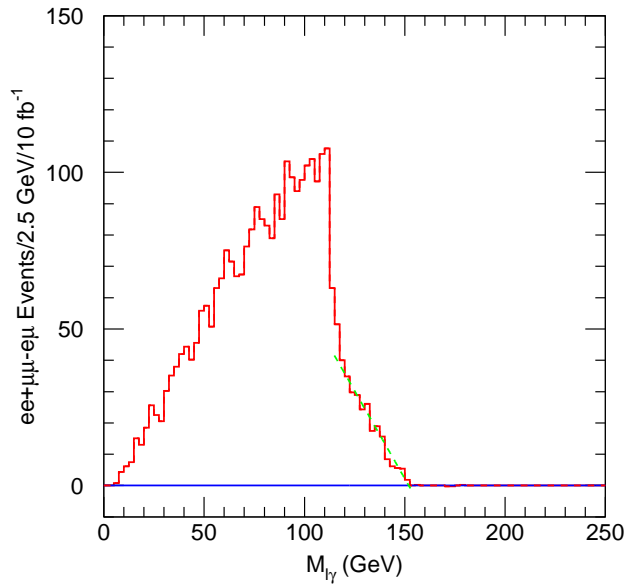}
\vskip-14pt
\caption{$M_{\ell^\pm\gamma}$ distribution for $e^+e^- + \mu^+\mu^- -
e^\pm\mu^\mp$ events sample with two photons; a linear fit from 115 to 
150 GeV is used to determine the endpoint for the photon and the
``wrong'' lepton and is shown as a dashed line.\label{mlg}} 
\end{figure}

The subset of events where one $\ell^+\ell^-\gamma$ mass is larger and
the other smaller than this endpoint was then selected. Only the
combination with the lower $\ell\ell\gamma$ mass can come from
$\tchi_2^0$ decay. The $M_{\ell^\pm\gamma}$ distribution for this
combination is shown in Figure~\ref{mlg}. Two structures are present
in this plot. There is a sharp edge at 
$$
\sqrt{M_{\tilde{\ell}_R}^2-M_{\chi_1^0}^2}=112.7\,\GeV
$$ 
from the photon and the second (``right'') lepton in the decay chain
($\tilde{\ell}\to \ell \gamma$),
and there is a distribution that vanishes linearly at
$$
\sqrt{M_{\chi_2^0}^2 -M_{\tilde{\ell}_R}^2}=152.6\,\GeV
$$ 
from the photon and the first (``wrong'') lepton. The former of these
has a background from the latter. The plot shows a linear fit between
115 and 150 GeV, which can be used to determine the position of the
second endpoint. We estimate the errors on these quantities to be $\pm
200$ and $\pm 500$~MeV for luminosity of 10~fb$^{-1}$ where they are
limited by statistics. We expect that that they will become limited by
systematics at $\pm 100$ and $\pm 200$~MeV at LHC design luminosity.  

These four measurements are sufficient to determine the masses of the
particles ($\tchi_2^0$, $\tell_R$, and $\lsp$) in this decay chain
without assuming any model of SUSY breaking. Of course the existence of,
and rate for, this decay chain are model dependent. So is the
interpretation of the slepton mass as the mass of the $\tell_R$. A
similar combination of three-body and four-body distributions will be
useful in other cases for which a decay chain involving three two-body
steps can be identified.\footnote{For example, at SUGRA Point 5
considered previously\cite{hinch,previous}, the decay $\tq_L \to
\tchi_2^0 q \to \tell_R^\pm \ell^\mp q \to \lsp \ell^+\ell^- q$ can be
used to determine an $\ell^+\ell^-$ edge, an $\ell q$ edge, and an
$\ell^+\ell^- q$ endpoint.}

\subsection{Reconstruction of $\tG$ Momenta}
\label{sec:recon}

The supersymmetry events each have two unobserved $\tG$'s. The sum of
their transverse momenta is, up to resolution effects and possible
missing neutrinos, measured as the two components of $\etmiss$.
There appears to be insufficient information to reconstruct their
momenta. However the decay chain
$$
\tchi_2^0 \to \tell^\pm\ell^\mp \to \lsp\ell^\pm\ell^\mp \to
\tG\gamma\ell^\pm\ell^\mp\,.
$$
provides three precisely measured final state particles and three mass
constraints using the masses determined from the measurements 
in the previous subsection. A $0C$ fit for the  $\tG$ momentum is
then possible assuming that $M_{\tG}=0$. The solution has a 4-fold
ambiguity
since the leptons cannot be uniquely assigned and 
there is quadratic ambiguity from the solution to the constraints.
The details  are presented in the appendix. 

\begin{figure}[t]
\dofig{3.5in}{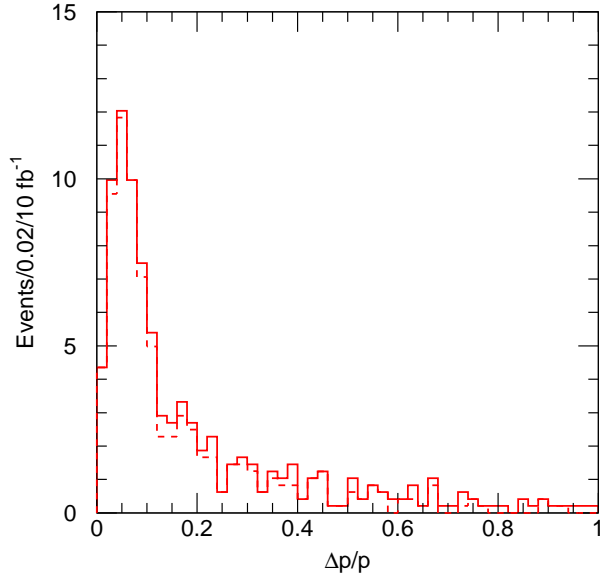}
\vskip-14pt
\caption{The distribution in $\abs{\Delta p}/\abs{p}$ for reconstructed 
 gravitino
momenta. The gravitino momenta are reconstructed using the 
method described in the text. The difference ($\Delta p$) 
between the reconstructed ($p$) and 
generated momentum is then formed for the combination with 
the lowest
$\chi^2$. Events are included where the lowest $\chi^2$ is 
less than 10 or 1 (dashed line).\label{dpp}}
\end{figure}

We want to have a $\tchi_2^0 \to \ell^+\ell^-\gamma\tG$ decay on both
sides of the event. We therefore select events with four leptons and
two photons. We require two opposite-sign, same-flavor lepton pairs
and a unique way of combining these with the photons consistent with
$\tchi_2^0$ decay. Then there are a total of 16 solutions. $\etmiss$
can be used to resolve these ambiguities as follows. The best
solution is selected by summing the momenta of the two $\tG$'s and
calculating the $\chi^2$ for matching this to the measured value of
$\etmiss$,
$$
\chi^2 =
\left({\slashchar{E}_x-p_{1x}-p_{2x}\over\Delta\slashchar{E}_x}\right)^2
+\left({\slashchar{E}_y-p_{1y}-p_{2y}\over\Delta\slashchar{E}_y}\right)^2\,.
$$ 
We then select the solution that has the lowest $\chi^2$
($\chi^2_{\rm min}$).  It is assumed that the
resolution in $\etmiss$ is determined by the total transverse energy
$E_T$,
$$
\Delta\slashchar{E}_x = \Delta\slashchar{E}_x = 0.6\sqrt{E_T} + 0.03 E_T
$$

We can evaluate the effectiveness of this method by comparing the
reconstructed values of the $\tG$ momentum to the best match with the
generated values. Figure~\ref{dpp} shows the distribution of the
fractional difference between the generated and reconstructed
 $\tilde{G}$ momentum
$$
{\Delta p \over p} = {\abs{\vec{p}_{\rm recon} - \vec{p}_{\rm gen}}
\over \abs{\vec{p}_{\rm gen}}}
$$
for solutions where $\chi^2_{\rm min} < 10$; the distribution for
$\chi^2_{\rm min}<1$ is very similar.  As can be seen from the figure
which indicates a resolution of order
10\% with a long tail, showing that the method works quite
well. In approximately 40\% of the events that enter this analysis,
the $\tilde{G}$ momenta reconstructed within 10\% of their nominal
value.
    The
shape of this distribution is dominated by detector resolution on the
leptons and photons. It is much narrower if the generated momenta are
used, and it is significantly wider if the resolution for the electron
energy is taken from the calorimeter alone; the central tracker is
important for soft electrons.\footnote{It is possible in principle to
apply this method to other cases involving three identifiable decays,
e.g., the $\tq_L$ decay chain for SUGRA Point~5\cite{hinch,previous}.
Unfortunately, the combination of more combinatorial background and
relatively poor resolution for jets seems to make it not possible to
reconstruct the $\lsp$ momenta in this case.}

\begin{figure}[t]
\dofig{3.5in}{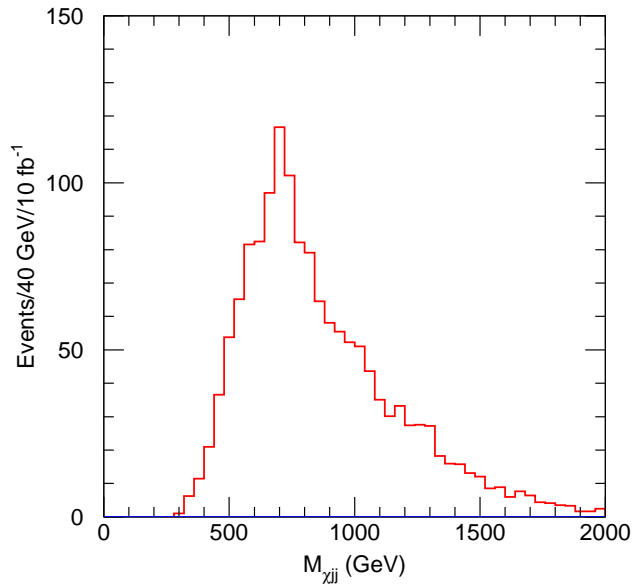}
\vskip-14pt
\caption{Invariant mass of  $\tchi_2^0$ and two jets, 
for events at Point G1a. The peak is due to the decay
 $\tilde{g} \to q\overline{q} \tchi_2^0$.  Events are required to have
 two $\tchi_2^0$ momenta  reconstructed and  at
least four jets with $p_T>75$ GeV.
\label{chi2jj}}
\end{figure}

\subsection{Reconstruction of Gluinos and Squarks}

\begin{figure}[t]
\dofig{3.5in}{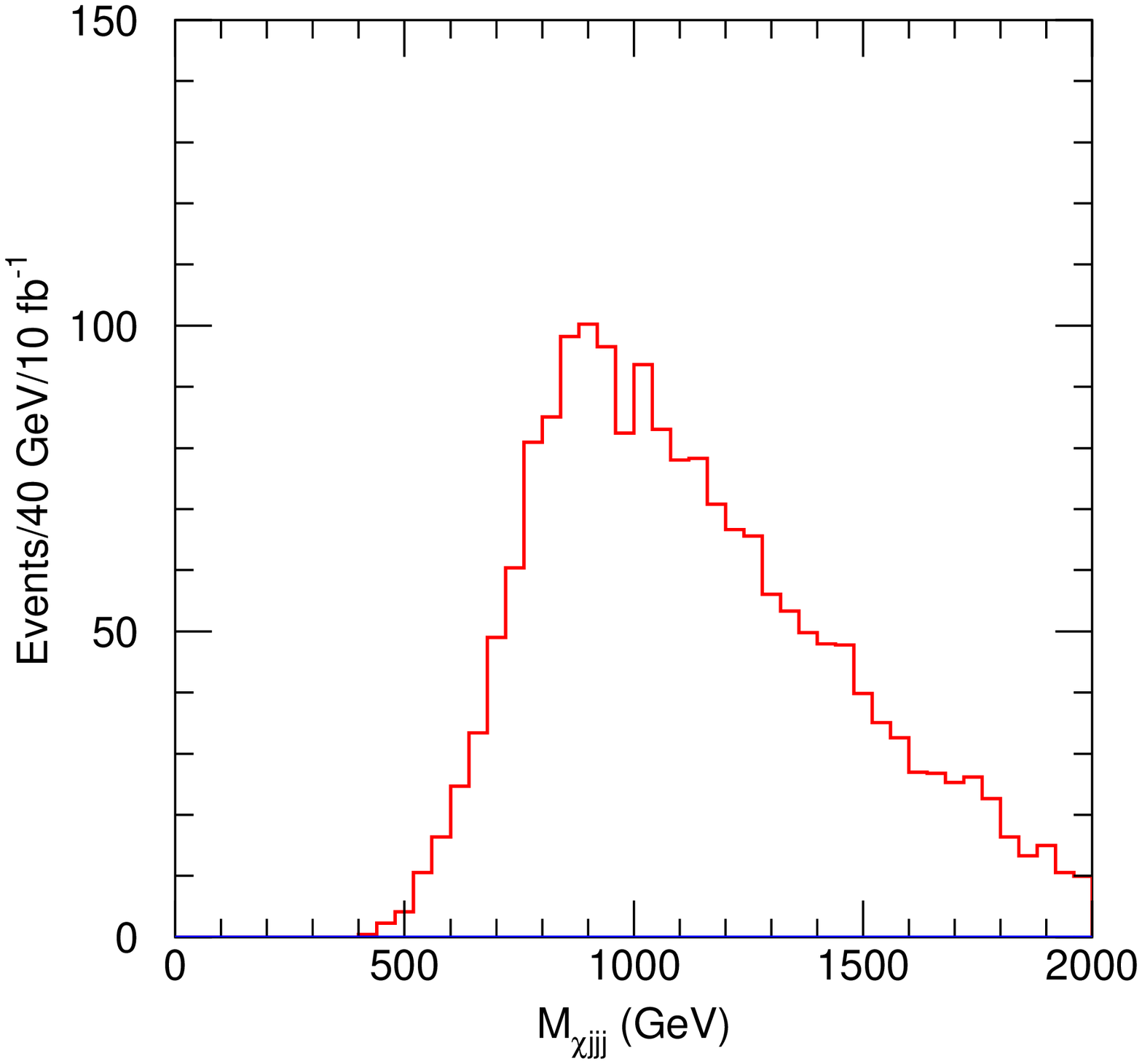}
\vskip-14pt
\caption{Invariant mass of  $\tchi_2^0$ and three jets,
 for events at Point G1a. The peak is due to the decay 
$\tilde{q}\to q\tilde{g}
 \to qq\overline{q} \tchi_2^0$.
\label{chi2jjj}}
\end{figure}

Squark production is significant at Point G1a 
 even though squarks are considerably
heavier than gluinos. The reconstruction technique of the 
previous section can be extended
to allow the chain $\tq \to \tg q \to \tchi_2^0 \overline{q}q q$ to be 
reconstructed and the gluino
and squark masses measured. We begin with the events selected in the
previous section where two $\tchi_2^0$
momenta each 
arising from $\tchi_2^0 \to \ell^+\ell^- \gamma \tG$ have been reconstructed.
Jets are are then searched for  that have
$p_T>75\,\GeV$ in a cone $R=0.4$. We require that at least four such jets
be present in the event.
Each $\tchi_2^0$
is then 
combined with two and with three jets; the resulting invariant mass 
distributions are shown in Figures~\ref{chi2jj} and \ref{chi2jjj}
respectively.
The $jj\tchi_2^0$ mass of
Figure~\ref{chi2jj} has a peak close to the gluino mass,
of $744\,\GeV$, while Figure~\ref{chi2jjj} has a broader peak near the
squark masses, $935$--$985\,\GeV$. The peaks occur in essentially
the same place if the jet cut is raised to $p_T>100\,\GeV$ and so are
{\it not} simply reflections of the kinematic cuts. It is important
to emphasize that this technique enables the masses of the squarks and 
leptons to be measured directly rather than being inferred from
features in kinematic distributions.

\begin{figure}[t]
\dofig{3.5in}{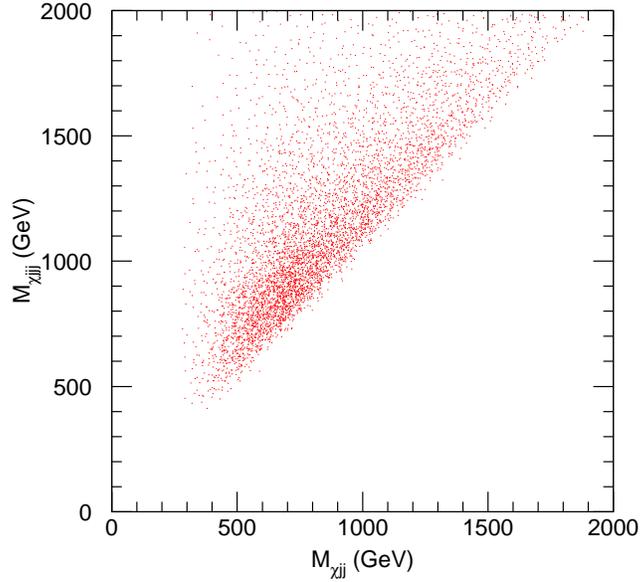}
\vskip-14pt
\caption{Scatter plot of $M(\tchi_2^0jjj)$ vs.{} $M(\tchi_2^0jj)$ for events
at Point G1a where two $\tchi_2^0$ momenta are reconstructed and there are at
least four jets with $p_T>75$ GeV.
\label{chi2scat}}
\end{figure}

\begin{figure}[t]
\dofig{3.5in}{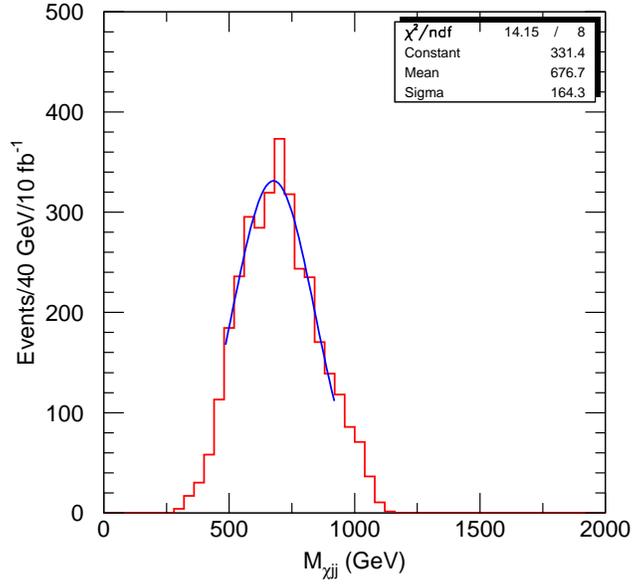}
\vskip-14pt
\caption{Projection of Figure~\protect\ref{chi2scat} on the
$M(\tchi_2^0jj)$ axis for $800\,\GeV < M(\tchi_2^0jjj) < 1000\,\GeV$.   The peak corresponds to gluino decay. The fit
shown is a Gaussian over the range 500 to 900 GeV.
\label{chi2jjproj}}
\end{figure}

\begin{figure}[t]
\dofig{3.5in}{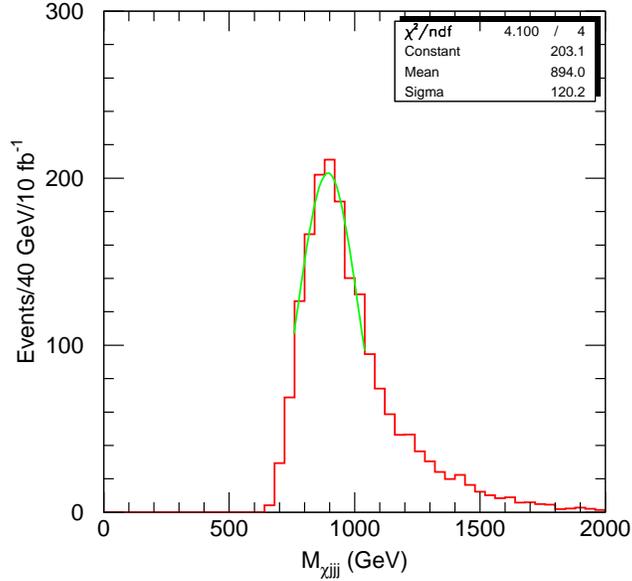}
\vskip-14pt
\caption{Projection of Figure~\protect\ref{chi2scat} on the
$M(\tchi_2^0jjj)$ axis for $600\,\GeV < M(\tchi_2^0jj) < 800\,\GeV$.
The peak corresponds to squark decay.  The fit
shown is a Gaussian over the range 760
 to 1020 GeV.\label{chi2jjjproj}}
\end{figure}

The peaks can be sharpened up considerably by searching for
correlations since, for each $\tchi_2^0$ momentum and set of
three jets, there should be
one $\tchi_2^0jjj$ peak at the squark mass mass and three
$\tchi_2^0jj$ mass combinations, one of which is at the gluino mass. 
The scatter
plot of all combinations is shown in Figure~\ref{chi2scat}. While the
points show significant scatter, there is, nevertheless, a clear peak in the
$M(\tchi_2^0jj)$ projection, made by selecting 
events in the range $800\,\GeV < M(\tchi_2^0jjj) < 1200\,\GeV$,
 shown in Figure~\ref{chi2jjproj}. The smooth curve
shown in this plot is a  Gaussian fit over the range 500 to 900 GeV;
 shows a clear
maximum at $699\,\GeV$. If a  cut
around this peak, $600\,\GeV < M(\tchi_2^0jj) < 800\,\GeV$, is made a 
projection of the scatter plot onto the $jjj\tchi_2^0$ axis made as
shown in
Figure~\ref{chi2jjjproj}, a somewhat narrower peak at the squark mass
than that of Figure~\ref{chi2jjj}
is obtained.
The smooth curve is again a Gaussian fit over the range 760 to 
1020 GeV and has a maximum at
about $909\,\GeV$.
The statistical errors on the locations of the peaks are quite
small and the precision of the measurements is likely dominated by systematic 
effects such as the calibration of the jet energy scale.

\section{Point G1b}
\label{sec:g1b}

In this case the NLSP is neutral and long lived.  Almost all of the
produced NLSP's exit the detector without interacting, so the
phenomenology is qualitatively similar to the SUGRA models. We first
see evidence for new physics via the presence of events with large
total $E_T$, large $\etmiss$, and isolated leptons.  Approximately
0.1\% of the NLSP's will decay within the detector volume, resulting
in a photon that does not point to the interaction vertex. The ability
to identify such photons and measure the decay point would provide a
valuable constraint on the lifetime and hence information on
$\Cgrav$. 
We anticipate
that a decay probability of 0.1\% will be difficult but perhaps
possible to detect.
Apart from this feature, this case is similar to one of the SUGRA cases
(Point 4) studied earlier \cite{hinch,fabiola}. In that case, there was 
more structure in the dilepton spectra as significant production of
$\tchi_3^0$
and $\tchi_4^0$ occurs in the decay of a gluino.

\label{sec:G1b}
\subsection{Effective Mass Analysis  at Point G1b}

\begin{figure}[t]
\dofig{3.5in}{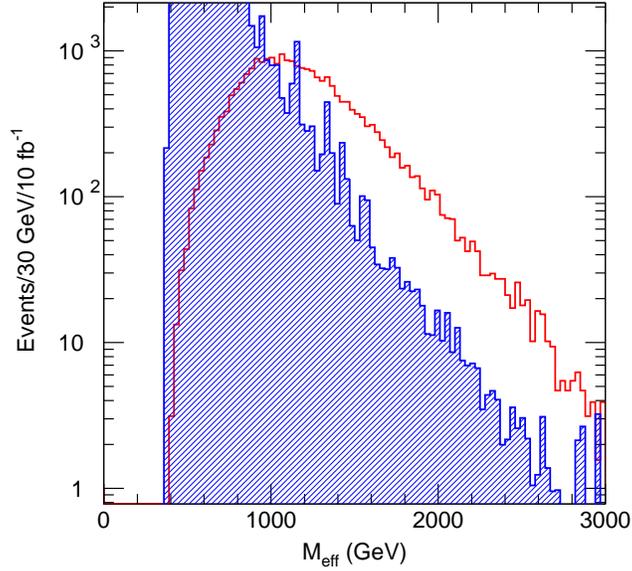}
\vskip-14pt
\caption{The effective mass distribution
 showing LHC Point~G1b signal and Standard Model backgrounds (hatched
 histogram) \label{G1befmass}}
\end{figure}

Events that have at least four jets are selected and the
scalar sum of the $p_T$'s of the four hardest jets and
the missing transverse energy $\etmiss$, formed
$$
\Meff = p_{T,1} + p_{T,2} + p_{T,3} + p_{T,4} + \etmiss\,.
$$
Here the jet $p_T$'s have been ordered such that $p_{T,1}$ is the
transverse momentum of the leading jet. Figure~\ref{G1befmass} shows
 the distribution in $M_{eff}$ for events where the following cuts
have been made. 
\begin{itemize}
\item   $\etmiss > 100\,\GeV$;
\item   $\ge 4$ jets with $p_T > 50\,\GeV$ and  $p_{T,1} > 100\,\GeV$;
\item   Transverse sphericity $S_T > 0.2$;
\item   No $\mu$ or isolated $e$ with $p_T > 20\,\GeV$ and $|\eta|<2.5$;
\item   $\etmiss > 0.2 \Meff$.
\end{itemize}
It can be seen clearly from this figure that there is an excess of events at 
large $\Meff$ over that expected in the Standard Model, providing clear
evidence for new physics.

\subsection{Selection of Dilepton events}

\begin{figure}[t]
\dofig{3.5in}{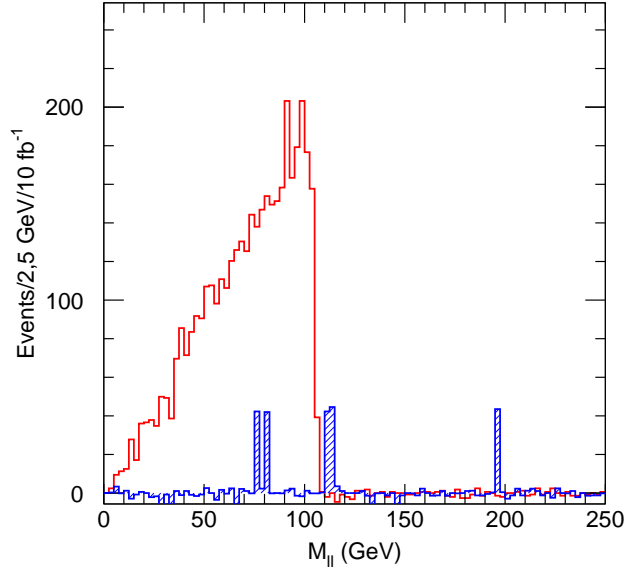}
\vskip-14pt
\caption{The dilepton mass distribution at Point~G1b. The Standard
  Model background (hatched histogram) is very small. The fluctuations
  in our event sample for the combination shown {\i.e.} $e^+e^- +
  \mu^+\mu^- - e^\pm\mu^\mp$ are therefore larger than one would
  expect in 10 fb$^{-1}$.
\label{g1bdilep}}
\end{figure}

We attempt to extract the decay chain $\tchi_2^0 \to \ell
\tilde{\ell}_R \to \ell^+\ell^- \lsp$ as follows.  Events are selected
that have $\Meff> 1000$ GeV and $\etmiss > 0.1\Meff$, and two and only
two isolated leptons of opposite charge with $p_T>20$~GeV for
electrons, $p_T>5$~GeV for muons, and $\abs{\eta}<2.5$ for both.  In
order to reduce the combinatorial background we again look at the
combination $e^+e^- + \mu^+\mu^- - e^\pm\mu^\mp$.
Figure~\ref{g1bdilep} shows the dilepton mass distribution. There is a
clear endpoint at
$$
M_{e}=M_{\tchi_2^0} \sqrt{1-\left(M_{\tell_R} \over M_{\tchi_2^0}\right)^2}
\sqrt{1-\left(M_{\lsp} \over M_{\tell_R}\right)^2} = 105.1\,\GeV.
$$
We expect this to have a precision of order 0.1\% on the position of
this end point. The Standard Model 
background shown on this plot reflects the poor
statistics of our sample, the actual background is much smaller.
The figure also shows a small, but statistically
significant, peak from $Z$ decays. These are arising from the decay
$\chi_2^0 \to \chi_1^0 Z$ which has a branching ratio of 9\%. The fact 
that this two body decay is of the same order of magnitude as the
$\ell^+\ell^-\lsp$ decay is strong evidence that the latter is
{\it not} arising from the three body decay 
$\chi_2^0 \to \chi_1^0 \ell^+\ell^-$ but from the sequence of two
2-body decays. Additional evidence is provided by the shape of the
edge. We can be confident therefore that the endpoint is measuring the
combination of masses in the above equation.

\subsection{Extraction of gluino}

\begin{figure}[t]
\dofig{3.5in}{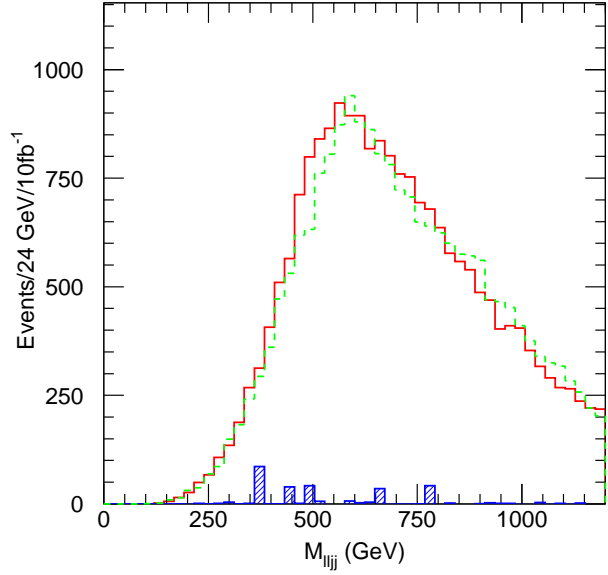}
\vskip-14pt
\caption{The mass distribution of a dilepton and two jets at Point
  ~G1b.
The dotted histogram corresponds to a different
  event sample where the gluino mass is changed to 800 GeV, this
  distribution has been scaled by a factor of 1.24 to facilitate
  comparison of the shapes. The Standard Model background is shown as
  the hatched histogram.
\label{G1bdilepjj}}
\end{figure}

We now select events from the previous sample that have at least two
jets with $p_T>125$ GeV and require that the  dilepton mass is
below 105 GeV.
The invariant mass of the 
$\ell^+\ell^- jj$ system is shown in Figure~\ref{G1bdilepjj} where 
again to reduce combinatorial background we look at  
the combination  $e^+e^- + \mu^+\mu^- - e^\pm\mu^\mp$. 
All combinations of jets with $p_T>25$~GeV are shown. The figure shows a 
broad peak but no clear structure. The events in this plot
are dominated by decay  $\tilde{g}\to q\overline{q} \tchi_2^0
\to 
q \overline{q}\ell^+\ell^- \lsp$. However the large combinatorial
problem prevents 
a kinematic edge from being visible. In order to estimate the
sensitivity to the gluino mass, we generate another set of events
which differ only in that the gluino mass has been increased to 800
GeV. The distribution from this sample is shown as the dashed
histogram in Figure~\ref{G1bdilepjj}. The total event rate for this
sample has been increased by a factor of 1.24, to compensate for the
smaller production rate and to facilitate a comparison of shapes of
the distributions. It is clear from this figure that these two curves
could be distinguished and that a constraint on the gluino mass
obtained.
The different event {\it rates} is not directly usable as a mass
constraint as the branching ratios are not known.

\begin{figure}[t]
\dofig{3.5in}{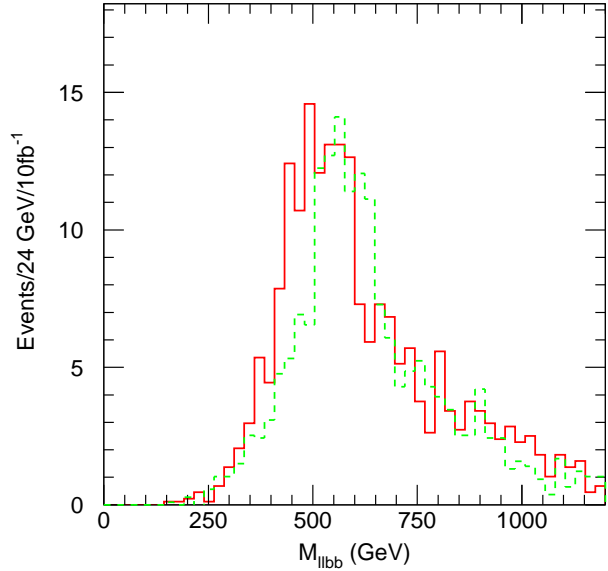}
\vskip-14pt
\caption{The mass distribution of a dilepton and two b-jets at Point
  ~G1b. 
The dashed histogram corresponds to a different
  event sample where the gluino mass is changed to 800 GeV. 
No standard model background events passed 
  the cuts.
\label{G1bdilepbb}}
\end{figure}

The signal can be improved somewhat if we restrict the jets to be
those that arise from bottom quark jets. The resulting distribution is 
shown in Figure~\ref{G1bdilepbb} which is identical to that for 
Figure~\ref{G1bdilepjj} except that the jets are required to be tagged 
as b-jets. The mass distribution has a clear structure which reflects
the kinematic end point of the dijet-dilepton system at
\begin{eqnarray*}
&&\vbigl[{50pt}(M_{\tilde{g}}-M_{\chi_2^0})^2+M_{e}^2 \\
&&\quad +\frac{M_{\tilde{g}}-M_{\tchi_2^0}}{2M_{\tchi_2^0}}
\vbigl({48pt}M_{\tilde{\tchi_2^0}}^2-M_{\tilde{\ell}}^2 +
\frac{(M_{\tchi_2^0}^2+M_{\ell}^2)(M_{\ell}^2-M_{\tchi_1^0}^2)}{2M_{\ell}^2}
\left(1+\frac{M_{\tchi_2^0}^2-M_{\ell}^2}{2M_{\tchi_2^0}M_{\ell}}\right)
\vbigr){48pt}\vbigr]{50pt}^{1/2}
\end{eqnarray*}
$M_e$ is the value of the dilepton edge defined above (the b quark
mass is neglected in this result). This
dilepton-dijet endpoint 
corresponds to 629 GeV. The dashed line on the figure shows the
result when the gluino mass is increased to 800 GeV; the kinematic
end point is now at 673 GeV. The event rates in this plot are rather
low and to obtain precision using this method 100 fb$^{-1}$ of
integrated luminosity will be needed.

\begin{figure}[t]
\dofig{3.5in}{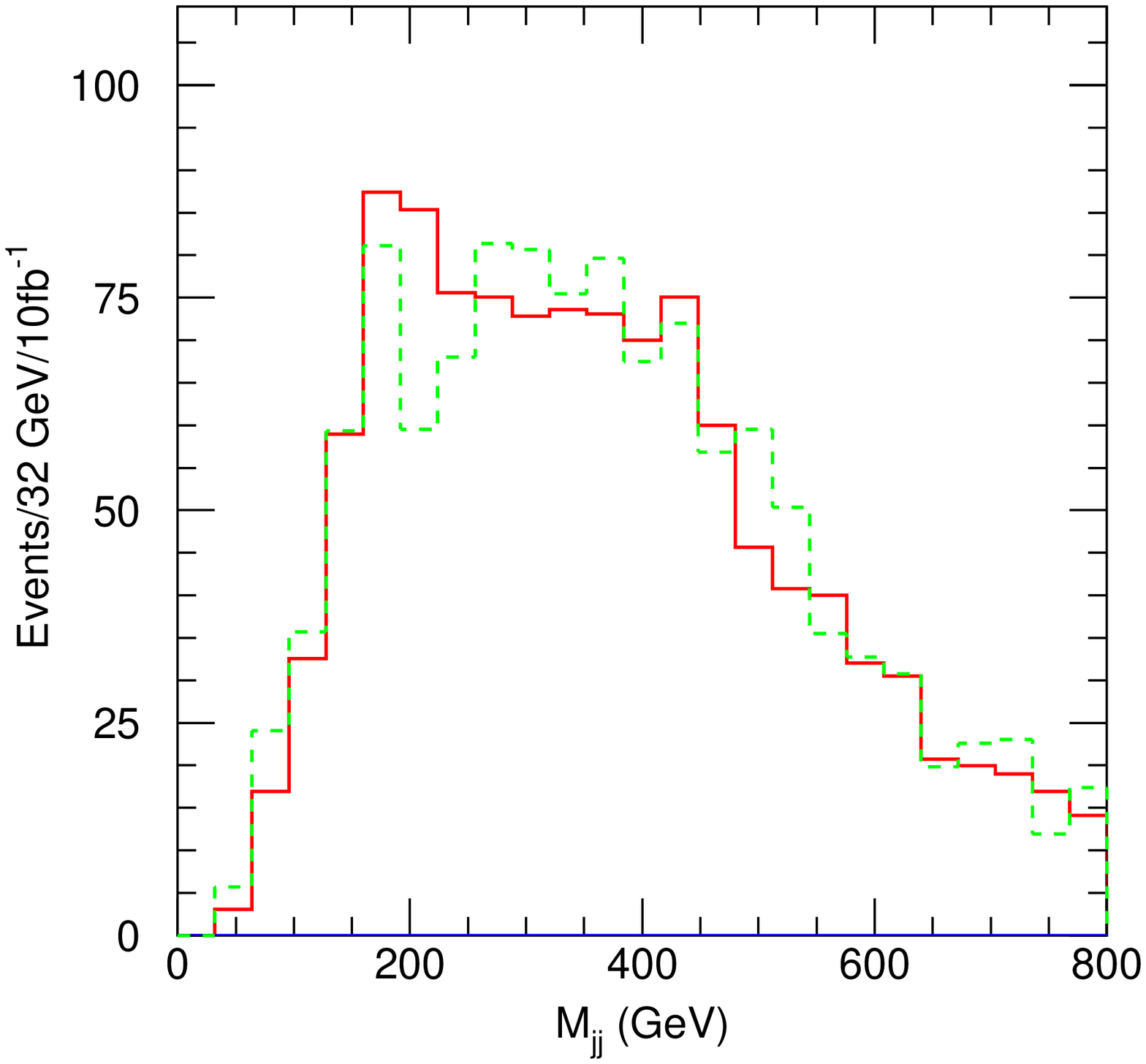}
\vskip-14pt
\caption{The mass distribution of two jets at Point ~G1b. There are two 
  entries per event.  The dotted histogram corresponds to a different
  event sample where the gluino mass is changed to 800 GeV, this
  distribution has been scaled by a factor of 1.18 to facilitate
  comparison of the shapes. No Standard Model events passed the cuts.
\label{G1bjj}}
\end{figure}

Following in the spirit of Ref.~\citenum{fabiola}, we attempt to look for
kinematic structure in the dijet mass distribution from the decay 
$\tilde{g}\to q\overline{q} \tchi_2^0 \to
q\overline{q} \ell^+\ell^- \lsp$. We begin by selecting events
with $\Meff> 1000$ GeV and $\etmiss > 0.1
\Meff$  and four
isolated leptons. We retain the event if these lepton can be grouped
into two opposite-sign, same  flavor pairs each of which has an
invariant mass below 105 GeV. We then select the four jets with the
highest transverse momenta, and pair them up, selecting the
combination that gives the smallest value for the sum of the two pair
masses.
A plot of the mass distribution of a dilepton pair and one of the jet
pairs is similar, with poorer statistics, to that of
Figure~\ref{G1bdilepjj}; there is still no clear kinematic feature. 
Figure~\ref{G1bjj} shows the invariant mass of each of the jet 
pairs(two entries per event). The
end-point that one expects to see at $m_{\tilde{g}}-m_{\tchi_2^0}$ is
not visible. However the shape of this curve is sensitive to the
gluino mass as can be seen by comparing the dashed histogram which is 
the result of the same event selection applied to an event sample
where the gluino mass is raised to 800 GeV. A detailed study
\cite{fabiola} of a case similar to this concluded that one could
measure the mass difference $m_{\tilde{g}}-m_{\tchi_2^0}$ with an
uncertainty of order 3\% with an event sample approximately 10 times
larger than this one (the cross-section is approximately 3 times
larger due to the smaller gluino mass of 580 GeV and 30 fb$^{-1}$ of
data was assumed). We will conservatively assume that we
can determine  $m_{\tilde{g}}-m_{\tchi_2^0}$ with an uncertainty of 
50~GeV, {\it i.e.} the difference shown in the figures.

\subsection{Extraction of squarks}

\begin{figure}[t]
\dofig{3.5in}{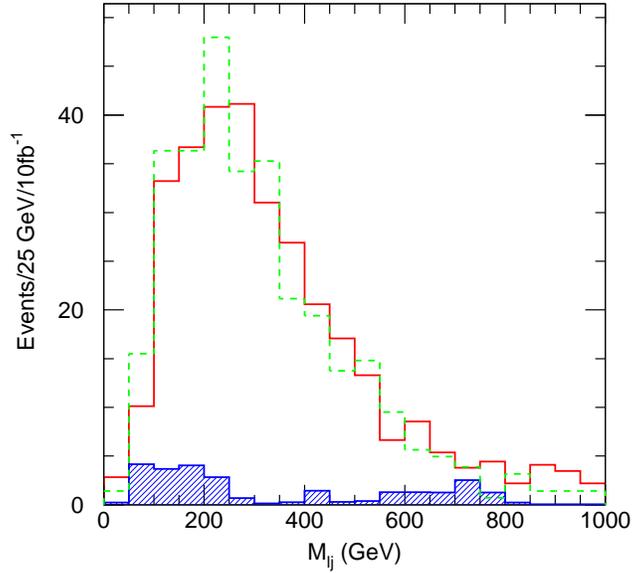}
\vskip-14pt
 \caption{The mass distribution of a lepton+ jet system. The dashed
 histogram corresponds to a different
  event sample where the squark mass has been reduced by  50 GeV. The
  Standard Model background shown as a hatched histogram is dominated
  by $t\overline{t}$ production.
\label{g1blj}}
\end{figure}

In view of the lack of clear structure in the events of the previous
section, it will be very difficult to extract the cascade decay
$\tilde{q}\to \tilde{g} q \to q\overline{q} 
q \tchi_2$  as was possible in the case of 
G1a. We therefore attempt to extract the decay $\tilde{q}\to q
\tilde{W} \to q W \lsp$.  We first select events with 
$\Meff> 1000$ GeV and $\etmiss > 0.1
\Meff$ and only two isolated leptons which can form any of the
following combinations: $e^+e^+$, $e^-e^-$, $\mu^+\mu^+$, $\mu^-\mu^-$ 
or $e\mu$ (any charges).
 In addition we require that there be two jets each with
$p_T>450$ GeV. Few events pass this selection, but in those that do
60\% of the directly produced supersymmetric particles are squarks.
Figure~\ref{g1blj} shows the lepton jet invariant mass distribution.
In order to assess the sensitivity of this distribution to the squark
mass, another event sample was produced where the squark masses were
reduced by 50 GeV. The distribution from this sample is shown as the
dashed histogram in Figure~\ref{g1blj}. The shapes are significantly
different; the shifting to the left of the dashed histogram is
symptomatic of the smaller mass of the squark. A Kolmogorov test
applied to the shape of the histograms in Figure~\ref{g1blj} using
finer binning and  data 
samples corresponding to 30 fb$^{-1}$ of integrated luminosity
indicates that they are distinct with a probability of 92\%. Sensitivity to squark masses at the 50 GeV level should therefore be 
possible.

\section{Point G2a}

The total supersymmetry supersymmetry cross-section for both  cases G2a and
G2b is 23~pb, larger than 
that of cases G1 because the squark masses are considerably lower. 
Since the slepton is the NLSP, almost all supersymmetry events contain 
a pair of sleptons. Most of these sleptons are produced in the 
decays $\tchi_2^0 \to \ell \tilde{\ell}$ and $\tchi_1^0 \to \ell \tilde{\ell}$.
In the case of Point G2a, the NLSP decays with a very short decay length
 (52 $\mu$) to give an additional lepton.
If the produced sleptons are selectrons or smuons, the high multiplicity
 of produced isolated 
leptons will provide both a convenient trigger and the first evidence for new 
physics. Only in the case where both sleptons are staus and decay 
hadronically, will we 
have to rely upon a jet and missing-$E_T$  trigger 
for the first indication of new physics.

\subsection{Dilepton Distributions}
\label{sec:G2alep}

As the branching ratios for $\tilde{\chi}_i^0\to \tilde{\ell}_R\to
\tilde{G}\ell^+\ell^-$ are substantial we can attempt to select this
decay chain by searching for events with isolated leptons and jets as
most of the $\tilde{\chi}_i^0$ will be produced in the decay of
strongly interacting sparticles. Events are selected that
have at least 4 jets with $p_T>25$ GeV and $\abs{\eta}< 2.5$
and at least 4 charged tracks associated with each jet
(to eliminate jets from hadronic tau decays). An $\Meff$ is formed
from the scalar sum of the transverse energies of the four jets with
the largest $p_T$ and $\etmiss$. 
$$
\Meff = p_{T,1} + p_{T,2} + p_{T,3} + p_{T,4} + \etmiss\,.
$$
We then require $ \Meff> 400 $ GeV and
$\etmiss >0.2 \Meff$. 

\begin{figure}[t]
\dofig{3.5in}{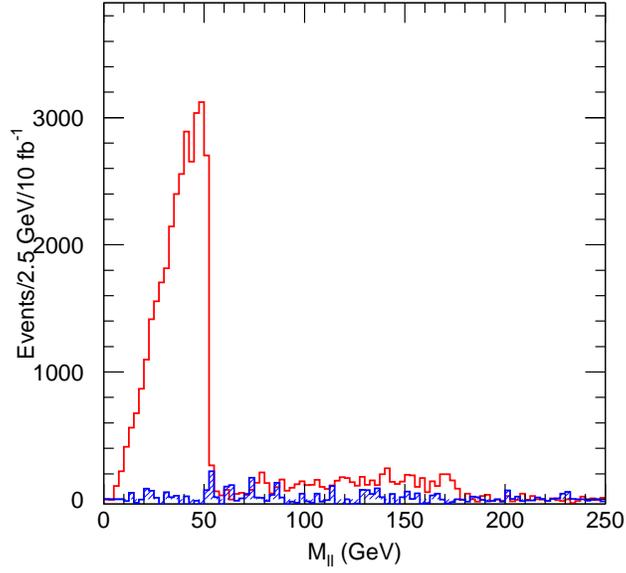}
\vskip-14pt
\caption{$M_{\ell\ell}$ distribution for the flavor subtracted
  combination
$e^+e^- + \mu^+\mu^- -
e^\pm\mu^\mp$ of events having two isolated leptons arising 
at Point G2a. The
Standard Model background shown does not represent the statistical
fluctuations expected in 10 fb$^{-1}$ of integrated luminosity (see text).
\label{alep}}
\end{figure}

\begin{figure}[t]
\dofig{3.5in}{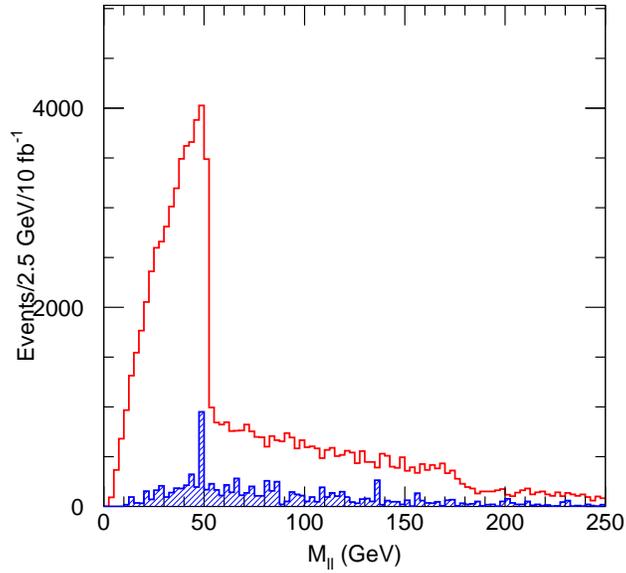}
\vskip-14pt
\caption{$M_{\ell\ell}$ distribution for all oppositely charged
  dilepton pairs $e^+e^- + \mu^+\mu^- +
e^\pm\mu^\mp$ in events having two isolated leptons arising 
at Point G2a. The background, mostly from $t\overline{t}$ production,
is shown as the hatched histogram.
\label{alepun}}
\end{figure}

An additional selection requiring two oppositely charged leptons
(either electrons or muons) with $p_T>10$ GeV and $\abs{\eta}<2.5$ is
made and the dilepton mass distribution formed. In order to reduce
combinatoric background, we form the combination 
$e^+e^-+\mu^+\mu^--e^\pm\mu^\mp$. The mass distribution for this
combination is shown in Figure~\ref{alep}. Two edges are visible at
52 and 175 GeV. corresponding to
$$
\sqrt{M_{\tchi_1^0}^2-M_{\tell_R}^2}=52.1\,\GeV
$$ 
and 
$$
\sqrt{M_{\tchi_2^0}^2-M_{\tell_R}^2}=175.9\,\GeV\,.
$$
The Standard Model background shown on this plot appears not to be 
 negligible.
However this is somewhat misleading as our background sample does not
correspond to the statistical fluctuations expected for an integrated
luminosity of 10 fb$^{-1}$. 
The number of background events can be seen more clearly in
Figure~\ref{alepun} which shows the combination
$e^+e^-+\mu^+\mu^-+e^\pm\mu^\mp$. It can be seen from this figure that 
this combination has considerable combinatorial background in the
signal events; the edge at 175 GeV is less clear. From this plot one can
estimate that, in  the
mass range 60 to 170 GeV, the  true Standard Model background in 
Figure~\ref{alep} is $0\pm 60$; this  fluctuation is insignificant 
compared with the $\sim 4000$ signal events in the same region.

\begin{figure}[t]
\dofig{3.5in}{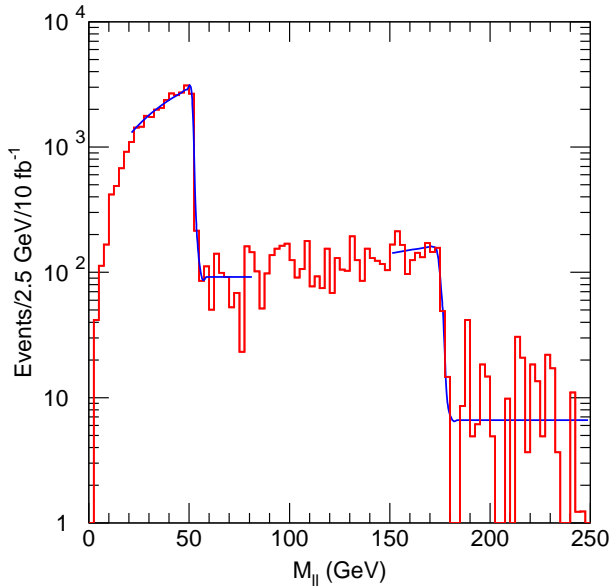}
\vskip-14pt
\caption{ As Figure~\protect\ref{alep} except that the distribution is
  shown on a logarithmic scale and the fits to the edges described in 
  the text are shown. No background is shown on this figure.
\label{G2bleplog}}
\end{figure}

It is clear that the statistical error on the precision the
measurement of the position of the lower
edge will be small and the systematic errors will dominate.
 The higher edge has much poorer statistics, so we have used a
fit to estimate how well the endpoint might be measured. Before any
cuts, the signal distribution in $M_{\ell\ell}$ has the form
$$
\frac{dN}{dM_{\ell\ell}}=A(1+M_{\ell\ell})\theta(M_{\rm max}-M_{\ell\ell})
$$
where $M_{\rm max}$ is the position of the edge. We add to this a
background $B$ taken to be constant in the vicinity of the edge and then
smear with a Gaussian of width $\sigma_M$ to be determined by the fit.
We then use the PAW version of MINUIT\cite{minos} to fit the data set of
Figure~\ref{alep}, whose statistical fluctuations represent
approximately $10$ fb$^{-1}$ of integrated luminosity, over the ranges
$20\,\GeV <M_{\ell\ell}< 80\,\GeV$ and $150\,\GeV <M_{\ell\ell}<
250\,\GeV$. The best fits are shown in Figure~\ref{G2bleplog}, and the
fits give

\begin{center}
\begin{tabular}{ll}
$M_{\rm max}=52.266^{+0.058}_{-0.045}$ GeV   & $\sigma_M=0.86$ GeV \\
\noalign{\smallskip}
$M_{\rm max}=175.46^{+0.21}_{-0.22}$  GeV  & $\sigma_M=6.5 $ GeV
\end{tabular}
\end{center}

\noindent The errors on $M_{\rm max}$ are determined by
MINOS\cite{minos}. We can expect systematic uncertainties of order
$0.1\%$ on these measurements.  We therefore expect that the precision
of the upper edge will be limited by statistics even for 100~fb$^{-1}$
of data. For the purposes of parameter fitting below, we will assume
that the errors are 70 and 270~MeV for 10~fb$^{-1}$ and 50 and 180~MeV
for 100~fb$^{-1}$ respectively.

\subsection{Detection of $\tq_R$}

\begin{figure}[t]
\dofig{3.5in}{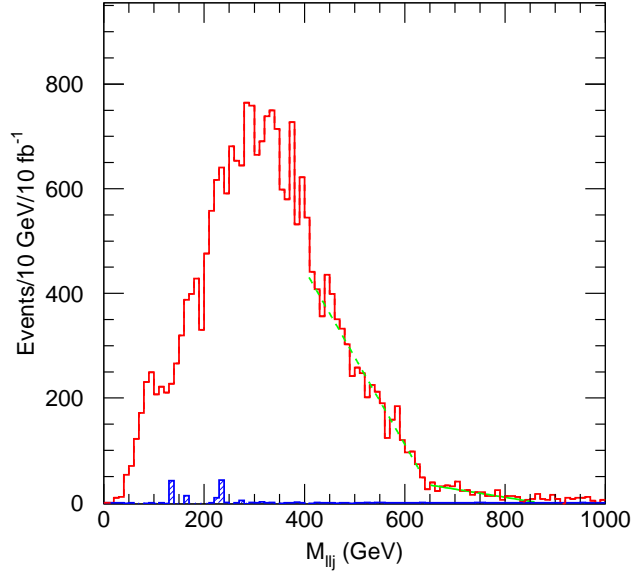}
\vskip-14pt
\caption{$M_{\ell\ell j}$ mass spectrum at Point G2a for events that
  contain at least two isolated leptons and four jets. The linear fit
  shown  as a dashed line 
is over the range 390 to 590 GeV. The Standard Model
  background is shown as the hatched histogram.
\label{stau1_mllq}}
\end{figure}

\begin{figure}[t]
\dofig{3.5in}{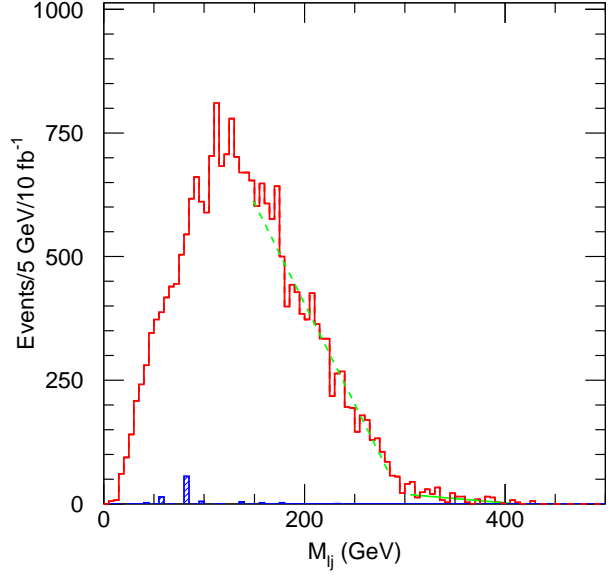}
\vskip-14pt
\caption{$M_{\ell j}$ mass spectrum at Point G2a. The  fits
  shown as dashed lines correspond to two separate linear fits over the ranges
is over the ranges 150  to 280 GeV and 305 to 400 GeV. The former
extrapolates to an endpoint at 315 GeV. The Standard Model
  background is shown as the hatched histogram.
 \label{stau1_mlq}}
\end{figure}

Here we use the decay chain $\tq_R \to q \tchi_1^0 \to q
\tell_R^\pm\ell^\mp \to q\tilde{G}\ell^+\ell^-$. The same event
selection as in Section~\ref{sec:G2alep} is used with the addition of
the  requirement that  the dilepton pair have mass $M_{\ell\ell}< 52$
GeV and 
transverse momentum $p_T(\ell\ell)>75$ GeV to enhance the probability that
the leptons come from the same decay chain.  The two jets with the
largest transverse momentum were then selected;
each is required to have 
$p_T>50\,\GeV$ and and  to contain
least 4 charged tracks. We show the invariant mass of the dilepton and
one of these two jets ($\ell^+\ell^- j$) in Figure~\ref{stau1_mllq} and 
of one the leptons and a jet in ($\ell^\pm j$) distributions are show in
Figure~\ref{stau1_mlq}. In each case only the smaller of the various
mass combinations is plotted and
 again we shown the subtracted distributions  
$e^+e^- + \mu^+\mu^- -e^\pm\mu^\mp$ which have a cleaner structure.

The $\ell^+\ell^- q$ distribution has an expected endpoint at 
$$
M_{\ell\ell q}^{\rm max} = \sqrt{M_{\tq_R}^2 - M_{\tell_R}^2} =
640.12\,\GeV
$$
while the $\ell^\pm q$ distribution has an expected endpoint at
$$
M_{\ell q}^{\rm max} = \sqrt{M_{\tq_R}^2 - M_{\lsp}^2} 
\sqrt{1- {M_{\tell_R}^2\over M_{\lsp}^2}} = 289.16\,\GeV\,.
$$
The plots show a linear fit below the end points. These fits 
extrapolate to end points slightly larger than the actual values.
The errors on
the precision of these values will be dominated by the systematic 
uncertainties in the jet energy scales estimated to be of order 1\%. 
We expect that the systematic 
uncertainty in the measurement of the {\it ratio} of these two
end-points will be less than this.

        We can now solve for the $\tell_R$, $\lsp$, and $\tq_R$ masses
in terms of the measured endpoints:
\begin{eqnarray*}
M_{\tell_R} &=& {M_{\ell\ell} \sqrt{M_{\ell\ell q}^2 -M_{\ell q}^2
-M_{\ell\ell}^2} \over M_{\ell q}}\\
M_{\lsp} &=& \sqrt{M_{\ell\ell}^2+M_{\tell_R}^2} \\
M_{\tq_R} &=& \sqrt{M_{\ell\ell q}^2+M_{\tell_R}^2}
\end{eqnarray*}
Note that this method for extracting masses requires only the existence
of the decay chain; the underlying model is not used in the analysis. Of
course the interpretation of the masses as those of $\tq_R$, $\tell_R$,
{\it etc.}, is model dependent.

\subsection{Detection of $\tilde{\chi}_1^\pm$  decays}

\begin{figure}[t]
\dofig{3.5in}{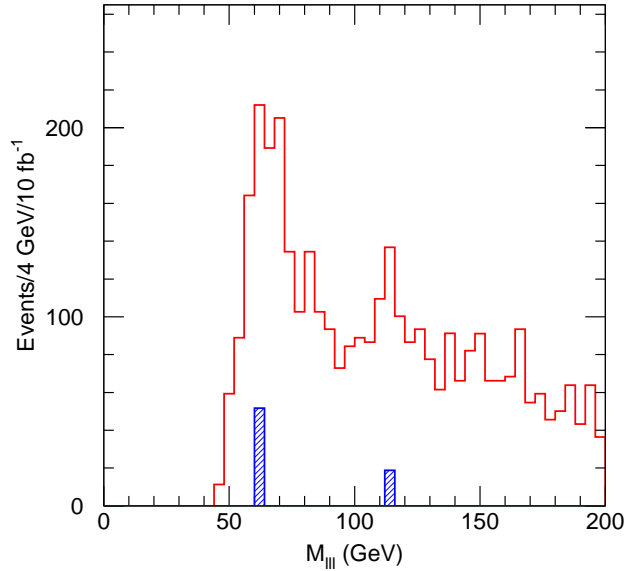}
\vskip-14pt
\caption{$M_{\ell\ell\ell}$ mass spectrum at Point G2a. The 
Standard Model background (hatched histogram) is very small.
 \label{tri-lep}}
\end{figure}

Approximately 50\% of the decays of $\tilde{q}_L$ occur to a 
$\tilde{\chi}_1^\pm$. There is a decay chain starting from 
$\tilde{\chi}_1^\pm$ proceeding through a slepton and a $\lsp$ that
gives a final state with three isolated leptons {\it viz.} 
$$\tilde{\chi}_1^+ \to \ell^+ \tilde{\nu} \to \lsp  \nu \ell^+ \to 
 \nu \ell^+\ell^-
\tilde{\ell}_R^+ 
 \to \nu \ell^+\ell^- \ell^+ \tilde{G}.$$
The combined branching ratio is 29\%.
We begin with the event selection of the previous section, requiring
that there be at least three isolated leptons in addition to the jets.
 Events are then
required to have at least one opposite sign, same flavor pair with an
invariant mass in the range $40 $ GeV $< M_{\ell\ell}<$ 52 GeV so that 
they are likely to have come from a $\lsp$ decay. Any other pairs of
leptons of same flavor  and opposite sign with  $ M_{\ell\ell}<$ 175 GeV 
are discarded as they are likely to come from $\tilde{\chi}_2^0$
decay. The selected dilepton is then combined with any other remaining 
lepton and the invariant mass of the trio is shown in
Figure~\ref{tri-lep}. If all three selected leptons arose from the
decay of $\tilde{\chi}_1^\pm$, this distribution would have a linear
vanishing at
$$
\sqrt{M_{\tilde{\chi}_1^\pm}^2-M_{\tilde{\nu}}^2+
M_{\lsp}^2-M_{\tilde{\ell}_R}^2}=85.75 \hbox{ GeV}
$$
There is considerable background in this plot. Nevertheless there is
clear evidence of structure. While it may not prove possible to
extract a precision measurement from this distribution, it provides
evidence for the existence of $\tilde{\chi}_1^\pm$ and a strong
consistency check of the model.

\section{Point G2b}

In this case the NLSP is the $\ttau_1$, and it has a long lifetime,
$c\tau \approx 1\,{\rm km}$.  The decay $\tell \to \ttau_1
\tau \ell$ is not kinematically allowed, so the $\tilde e_R$ and
$\tilde\mu_R$ are also long-lived. Each event will contain two of these
quasi-stable particles which will appear in the detector as a pair of
slow muons. These will provide a trigger as the mean velocity is quite
large (see below) and the first evidence for new physics in this case.

A small fraction of the sleptons will decay within the detector;
significant energy is released in the decay so the result is a track
which ends somewhere and a decay product that points back to this end
point. If the sparticle in question is a selectron, the resulting
electron could  be pointed back using the electromagnetic calorimeter
in combination with information
from the central tracker. If the sparticle is a smuon, the resulting
muon would have to be pointed using the remainder of the tracking volume
and the outer muon system. If it is a stau, the resulting hadronic decay
of the tau might be pointed using a combination of the central tracker
and the electromagnetic calorimeter.\footnote{To our knowledge no
detailed study of the smuon and stau cases has been done.} It is worth
emphasizing the importance of measuring this decay length: it is the
only way to obtain information on the gravitino couplings and the
fundamental scale of supersymmetry breaking.

\subsection{Effective mass analysis at Point G2b}

\begin{figure}[t]
\dofig{3.5in}{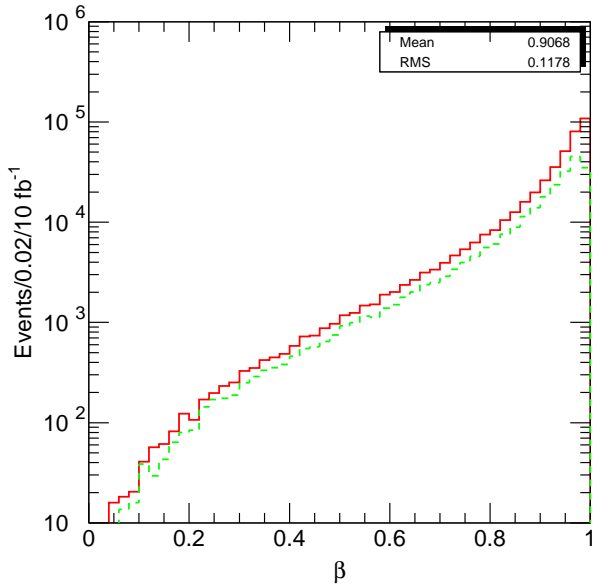}
\vskip-14pt
\caption{Generated slepton velocity distribution $\beta$ at Point G2b.
The dotted curve is for sleptons with $\eta<1$. 
\label{beta}}
\end{figure}

The events can be triggered using the quasi-stable particles that will
appear as muons. The velocity distribution of these particles is shown
in Figure~\ref{beta}, from which it can be seen that the mean velocity
is greater than $0.9c$.  Hence many of these should pass the ATLAS
level-1 muon trigger\cite{aleandro}.

\begin{figure}[t]
\dofig{3.5in}{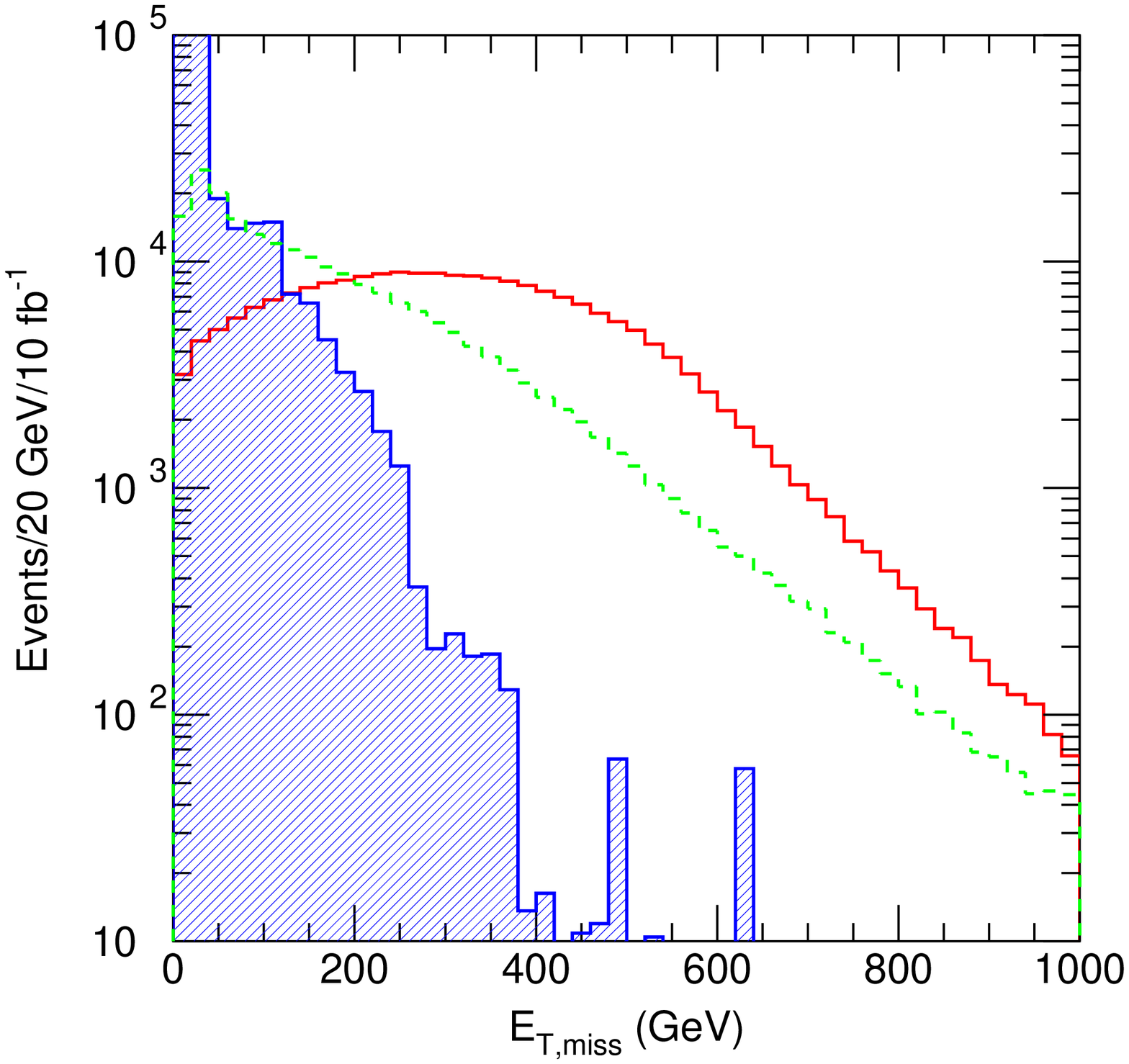}
\vskip-14pt
\caption{Calorimetric missing transverse energy $\etmiss$ at Point G2b.
The dotted curve shows the true $\etmiss$ including sleptons. The
Standard Model background of dimuon events is shown as the hatched
histogram.  \label{etmiss}}
\end{figure}

The events also have a large amount of missing $E_T$ as measured by
the calorimeter. The distribution is shown in Figure~\ref{etmiss},
which has a mean value of $315\,\GeV$. If the measured momenta of the
sleptons is included, the missing energy is much smaller as can be
seen from the dotted curve in Figure~\ref{etmiss}. The true missing
$E_T$ is larger than standard model backgrounds due to the larger
number of taus and heavy flavors in the SUSY sample. The Standard Model
background shown on this plot is controlled by the requirement that it 
contain  at least two muons. This calorimetric
missing $E_T$ could also be used as a trigger.

\begin{figure}[t]
\dofig{3.5in}{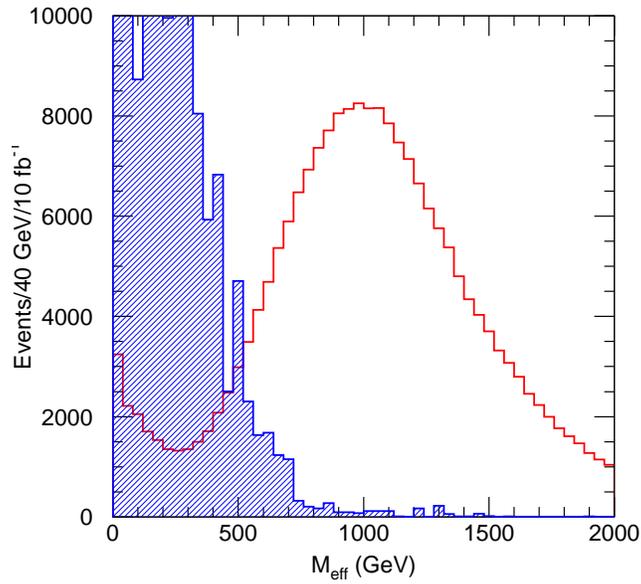}
\vskip-14pt
\caption{Effective mass distribution not including sleptons at Point
G2b.  The Standard Model
background of dimuon events is shown as the hatched histogram.
\label{meff}}
\end{figure}

We begin the analysis by using  
the effective mass distribution found by selecting
events that have at least four jets and forming
scalar sum of the $p_T$'s of the four hardest jets and
the missing transverse energy $\etmiss$,
$$
\Meff = p_{T,1} + p_{T,2} + p_{T,3} + p_{T,4} + \etmiss\,.
$$
Here the jet $p_T$'s have been ordered such that $p_{T,1}$ is the
transverse momentum of the leading jet. There is no requirement 
that the jets or missing $E_T$ be large enough to provide a trigger; we assume 
that the events are triggered by the muon system. Note that the
Standard Model
background shown on this plot is  required to have two muons in it and 
is suppressed as a result.
 The NLSP's are ignored when
making this effective mass variable. 
The distribution shown in Figure~\ref{meff}, 
has a mean value of $1004\,\GeV$ characteristic of the masses of the
strongly interacting sparticles which dominate the production.

\begin{figure}[t]
\dofig{3.5in}{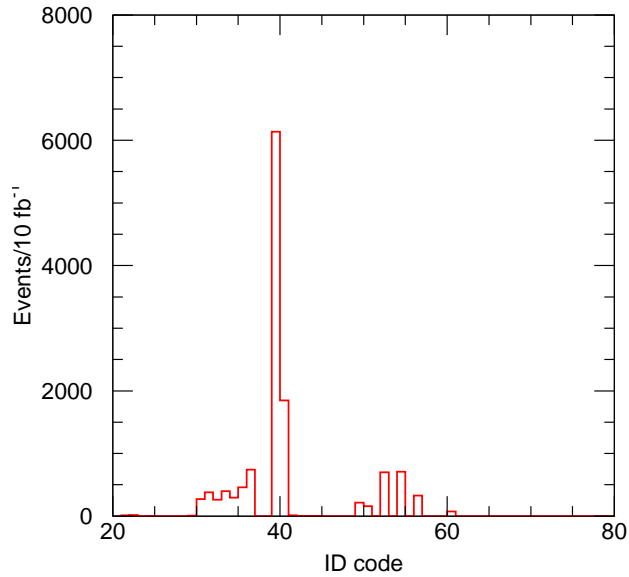}
\vskip-14pt
\caption{The particle ID codes for the primary particles of events in
Figure~\ref{meff} which have $\Meff< 100$~GeV: left sleptons (31-36),
gauginos (30, 39 and 40) and right sleptons (52, 54, and 56). 
\label{ident}}
\end{figure}

There is a peak in the effective mass distribution at zero,
corresponding to the production of events which have little hadronic
activity and $\etmiss$. It is due to the direct production of NSLP via
such processes as $q\overline{q}\to \tilde{\ell}^+\tilde{\ell}^-$ and
the direct production of $\tchi_1^0$ and $\tchi_1^+$ which then decay
to the NLSP. This is shown in Figure~\ref{ident} which shows the ID codes 
for the produced primary sparticles and demonstrates that they are
mainly sleptons and gauginos. We will
discuss these events in more detail below.

\subsection{Slepton mass determination}

The sleptons (NLSP) are dominantly produced at the end of decay chains 
and consequently the majority of them  are fast. 
The velocity distribution is shown in Figure~\ref{beta} from which the mean
velocity   can be determined.
Time
of flight measurements in the muon detector system can be used to
determine this velocity. When this is combined with a measurement of
momentum in the same system, the mass can be obtained.

\begin{figure}[t]
\dofig{3.5in}{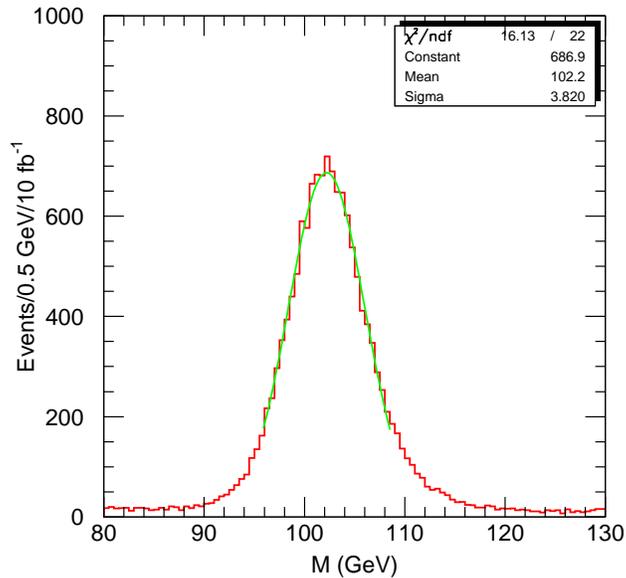}
\vskip-14pt
\caption{Reconstructed slepton masses for time delays $10\,\ns <
\Delta t < 50\,\ns$ relative to a $\beta=1$ particle. \label{mslep}}
\end{figure}

The muon chambers in the ATLAS detector can provide a time-of-flight
resolution of about $1\,\ns$.
For each slepton with $\abs{\eta}<2.5$ the time delay relative to $\beta=1$
to the outer layer of the muon system, taken to be a cylinder with a
radius of $10\,{\rm m}$ and a half-length of $20\,{\rm m}$, is
calculated using the generated momentum and smeared with a Gaussian
$1\,\ns$ resolution. The smeared time delay $\Delta t$ and measured
momentum
are then use to calculate a mass. The resulting mass distribution is
shown in Figure~\ref{mslep} for sleptons with $10\,\ns < \Delta t <
50\,\ns$. Raising the lower limit on $\Delta t$ improves the mass
resolution but reduces the efficiency. The resolution is never good
enough to resolve the $\ttau_1$ and $\ell_R$ masses of $101.35\,\GeV$
and $102.67\,\GeV$. 
The upper limit on $\Delta t$ is somewhat arbitrary; it reflects
practical concerns and also eliminates sleptons with very low $\beta$
that lose most of their energy in the calorimeter. The average of the
generated distribution, $102.2\,\GeV$, agrees well with the fitted mean
value in Figure~\ref{mslep}. It is important to note that 
this method will provide an mass measurement of an average over the 
$\ttau_1$, $\tilde{\mu}_R$ and $\tilde{e}_R$ masses as this analysis
cannot distinguish slepton flavors.

\subsection{Reconstruction of $\tchi_1^0$, $\tchi_2^0$ and $\tchi_4^0$}

\begin{figure}[t]
\dofig{3.5in}{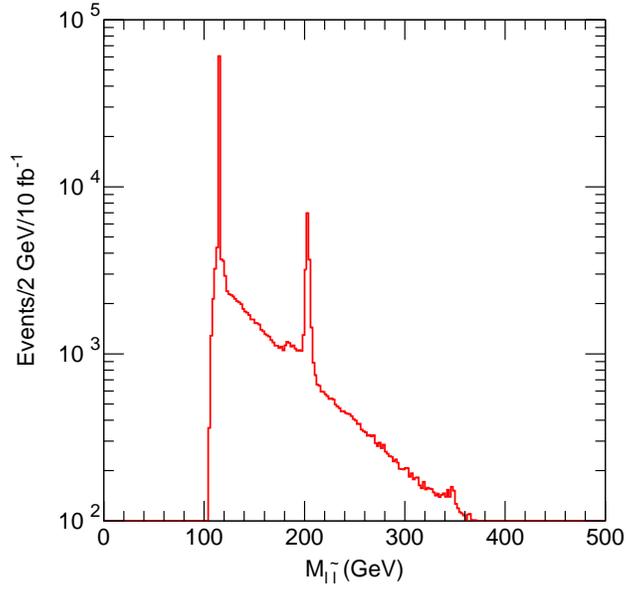}
\vskip-14pt
\caption{$\tell^\pm \ell^\mp$ mass distribution at Point G2b. \label{mchi}}
\end{figure}

        Since the $\tell_R$ are quasi-stable, the decays $\tchi_i^0 \to
\tell_R \ell$ can be fully reconstructed.  Events are selected that
have at least three isolated electrons, muons or quasi-stable sleptons
with $\abs{\eta}<2.5$ and $p_T>10$ GeV.  The two highest $p_T$
particles among the sleptons and muons ({\it i.e.} those particles that
penetrate to the muon system) are assumed to be sleptons and are
assigned their measured $\vec p$ and the slepton mass; the rest are
assumed to be muons.  The Standard Model background is already
negligible, so there is no need to make a time-of-flight cut to
identify the sleptons.  Sleptons are then combined with electrons or
muons ($\ell$) of the opposite charge and the resulting mass for all $\tell^\pm
\ell^\mp$ combinations is plotted in Figure~\ref{mchi}; we have no
way of determining the flavor of a slepton. There are two narrow peaks
at the $\lsp$ and $\tchi_2^0$ masses. The rather strange shape of the
$\lsp$ peak is a consequence of the fact that the splitting between
the $\lsp$ and $\tell_R$ is small, so the mass is dominated by the
rest mass of the $\tell_R$.
There is a small peak at 348 GeV due to
the decay of $\tchi_4$.     

\begin{figure}[t]
\dofig{3.5in}{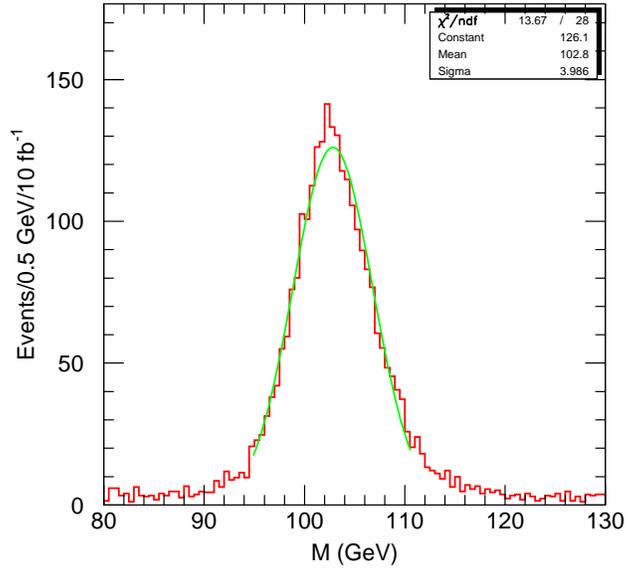}
\caption{Same as Figure~\protect\ref{mslep} for events in a $\pm 5$
  GeV
window around the $\lsp$
mass peak of Figure~\protect\ref{mchi} and the lepton is identified as 
an electron.\label{mell}}
\end{figure}

        The mass measurement of the slepton itself can now be refined 
and the smuon and selectron separated by
using the events in the
in the $\lsp$ mass peak. By restricting the event
samples to the cases where the {\it lepton} is either an electron or muon, 
this method can be used to provide separate samples of  
$\tilde{e}_R$ and $\tilde{\mu}_R$. The analysis of the previous section is
now repeated for the events within $\pm5\,\GeV$ of
the $\lsp$ peak where the lepton is an
electron.
The resulting distribution, shown in
Figure~\ref{mell}  has a
mean quite close to the correct $\tilde{e}_R$ mass. 
The statistical error on the mass in 
from a data sample of 10 fb$^{-1}$ is $\sim 100$~MeV.
Therefore it should be possible to  distinguish
the average slepton mass as determined in the previous section from the 
$\tilde{e}_R$  and  $\tilde{\mu}_R$ masses. 
The actual errors are likely to be dominated by 
systematic effects estimate to be 
0.1\%.
 This will be sufficient to constrain the mass of the stau
which is present in the average from the separate selectron  and smuon masses.

\subsection{Extraction of $\tilde{\ell}_L$}
\label{sec:slepton}

\begin{figure}[t]
\dofig{3.5in}{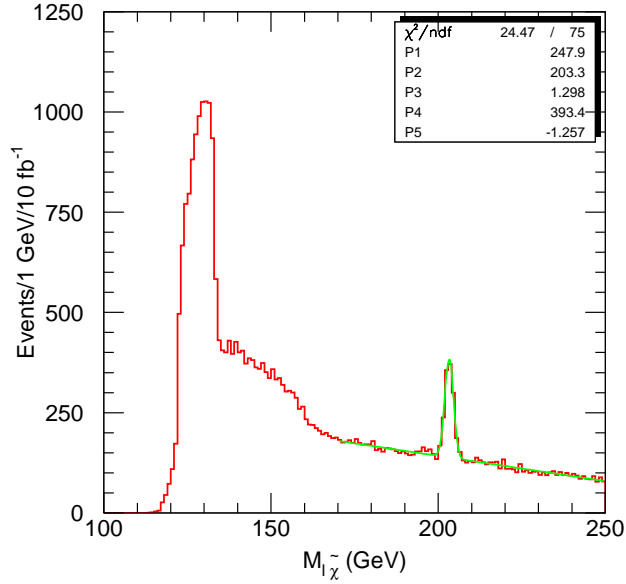}
\vskip-14pt
\caption{The mass distribution of a slepton and a pair of leptons at
  Point G2b. Events are
  selected so that a lepton and a slepton reconstruct to the $\lsp$
  mass ($M_\lsp \pm 5$ GeV). 
\label{mchil3}}
\end{figure}

We can begin with the reconstructed $\lsp$ and combine it with another
charged lepton in an attempt to detect the decay chain
$\tilde{\ell}_L\to \lsp \ell \to \tilde{\ell}_R^\pm\ell^\mp \ell$.
We select events that have at least two muons or stable sleptons with
$p_T> 10 $ GeV and $\abs{\eta} <
2.5$ The two highest $p_T$ objects are assigned to be sleptons and the 
rest are called muons. Combinations of a slepton and either muon or
electron
are formed that have no net charge and the system is tagged as
$\lsp$ if $M_{\ell^\pm\ell^\mp}= M_{\lsp} \pm 5$ GeV. This $\lsp$
candidate is then combined with another charged lepton and the mass of 
the three lepton system is shown in Figure~\ref{mchil3}.
There is a clear peak at the slepton mass of 203~GeV.

Another feature is also present in this
plot. The decay chain $\tilde{\chi}_1^\pm\to \tilde{\nu}\ell \to \lsp \nu 
\ell \to \tilde{\ell}_R^\pm\ell^\mp \nu\ell$
has a kinematic upper bound for the mass of the 
$\tilde{\ell}_R^\pm\ell^\mp \ell$ system of 
$$\sqrt{M_{\tilde{\chi}_1^\pm}^2 -M_{\tilde{\nu}}^2+M_{\lsp}^2 }=134 
\hbox{ GeV}
$$
The structure below this end-point is clearly visible in 
Figure~\ref{mchil3}. However there is a large background 
so a very accurate measurement will be difficult. Nevertheless, this
structure
measures a combination of $\tilde{\nu}$ and $\tilde{\chi}_1^\pm$  and
will provide a powerful constraint on the model.

\subsection{Reconstruction of Squarks} 
 
\begin{figure}[t]
\dofig{3.5in}{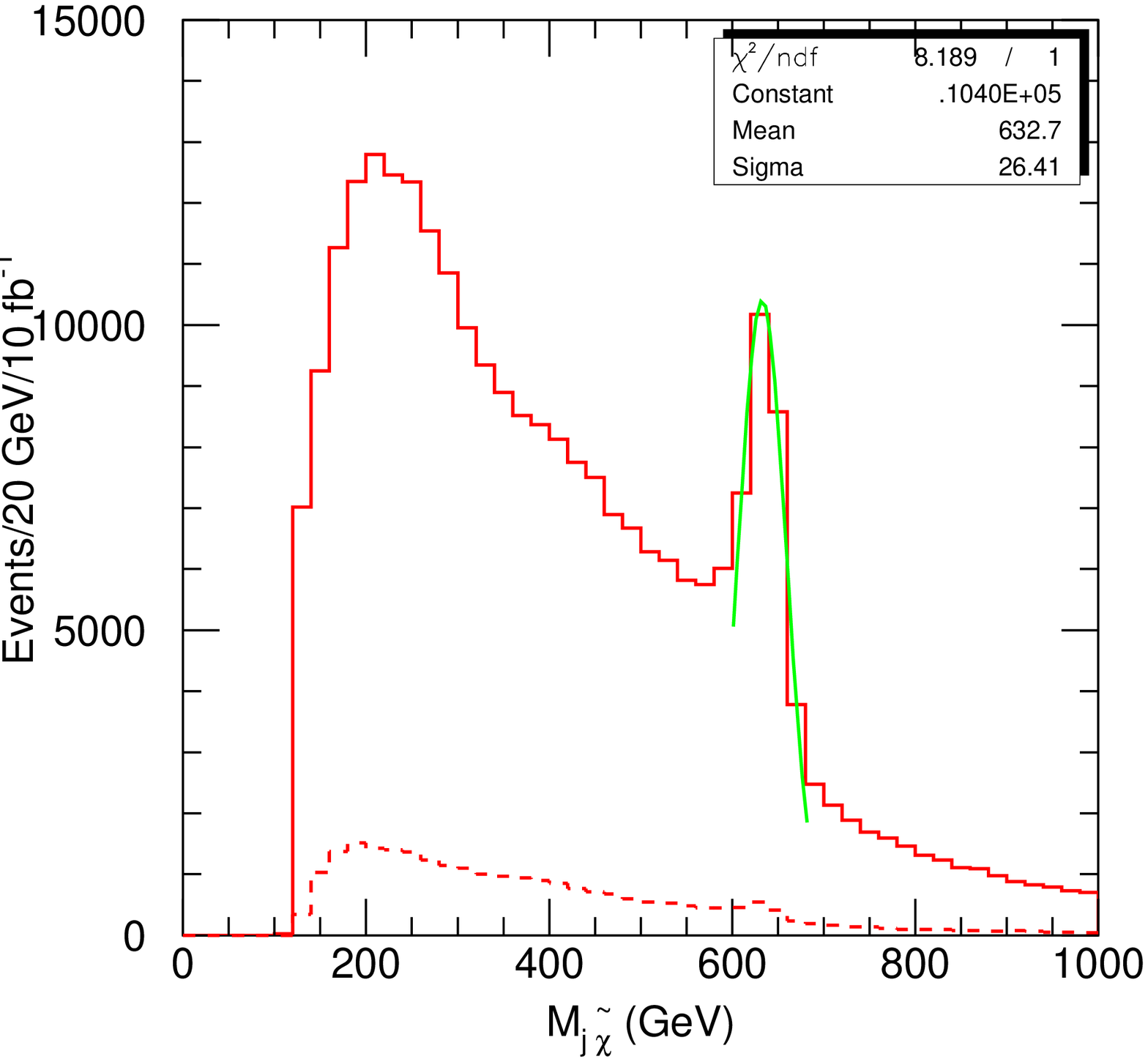}
\vskip-14pt
\caption{$\lsp$-$q$ mass distribution at Point G2b. 
Events  have a lepton and a slepton  with invariant mass in a $\pm 5$
  GeV
window around the $\lsp$
mass peak of Figure~\protect\ref{mchi}.
The dashed distribution
corresponds to cases where the jet is from a $b$-quark. The fit is a 
Gaussian over the range 610 to 640 GeV. \label{chi1q}}
\end{figure}

At this point squarks are considerably lighter than gluinos,
reflecting the fact that $m_{\tilde{g}}/m_{\tilde{q}}\sim \sqrt{N_5}$
and $N_5>1$, as is usually the case when the NLSP is a slepton. 
 Direct production of squarks
is dominant and the decay $\tq_R \to \tilde{\chi}_1^0 q$ is large.
Events are selected that have a $\ell\tilde{\ell}$ mass within 5 GeV
of the $\lsp$ peak in Figure~\ref{mchi}.  This pair is then combined
with any of the four highest $p_T$ jets in the event. The resulting
mass distribution of the jet-$\lsp$ system shown in Figure~\ref{chi1q}
has a relatively narrow peak somewhat below the average $\tq_R$ mass
of $648\,\GeV$. Some shift is expected since the jets are defined
using a small cone size, $R=0.4$.  

\begin{figure}[t]
\dofig{3.5in}{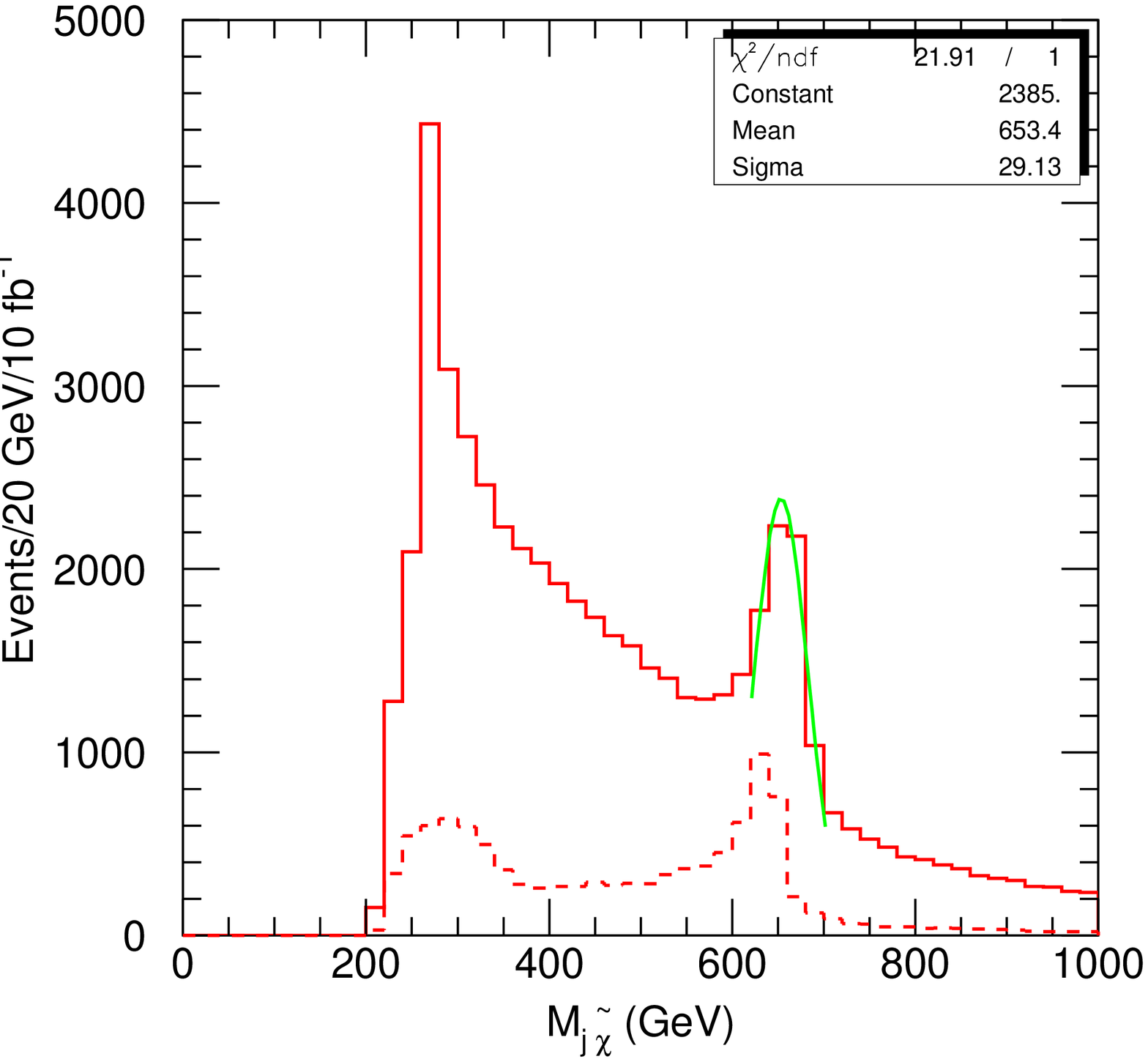}
\vskip-14pt
\caption{$\tchi_2^0$-$q$ mass distribution at point G2b. 
Events  have a lepton and a slepton  with invariant mass in a $\pm 5$~GeV
window around the $\tchi_2^0$. The dashed distribution
corresponds to cases where the jet is from a $b$-quark.The fit is a 
Gaussian over the range 625 to 700 GeV  \label{chi2q}}
\end{figure}

The decay $\tq_L \to \tchi_2^0 q$ is significant but not dominant.
It can be reconstructed by selecting events that have a
$\ell\tilde{\ell}$ mass within 5 GeV of the $\tchi_2^0$ peak in
Figure~\ref{mchi} and combining the $\tchi_2^0$ momentum with the
momentum of any of the four highest $p_T$ jets. The resulting mass
distribution of the jet-$\tchi_2^0$ system is shown in 
Figure~\ref{chi2q}; it has a somewhat wider peak than that of
Figure~\ref{chi1q} a little below the average $\tq_L$ mass of
$674\,\GeV$. The signal to background ratio is poorer than in
Figure~\ref{chi1q}, a reflection of the smaller branching ratio $\tq_L
\to \tchi_2^0 q$ ($\sim 25\%$). 

It is also possible to reconstruct the $\tilde b_{1,2}$ squarks, The
subset of events in Figure~\ref{chi2q} for which the jet is tagged as
a $b$ jet is shown in Figure~\ref{chi2q} as the dashed curve.  No
correction of the $b$-jet energy is done. The $\tilde{b}$ squark peak
is clearly visible and is at somewhat lower masses. The resolution is
insufficient to separate the peaks from $b_1$ and $b_2$ whose average
mass is 647 GeV and separation is 9 GeV.
  A subset of the events of Figure~\ref{chi1q} where
the quark jet is tagged as a $b$-jet is shown as the dashed line in
Figure~\ref{chi1q}.  No structure is visible in this case.
 This
 difference
  is due
 to the  branching ratios; $BR(\tilde{b_1}\to \lsp b)=5.9\%$,
  $BR(\tilde{b_2}\to \lsp b)=3.6\%$, $BR(\tilde{q_R}\to \lsp q)=94\%$, while
  $BR(\tilde{b_1}\to \tchi_2 b)=15.6\%$, $BR(\tilde{b_2}\to \tchi_2
  b)=10.8\%$, and $BR(\tilde{q_l}\to \tchi_2 q)=25\%$. The mixing
  between the b-squarks ensures that both can decay to $\tchi_2$. 

The wider peak of Figure~\ref{chi2q} relative to that of
Figure~\ref{chi1q} can now be understood. It is due to the presence of
a significant number of $b$-squarks in the former distribution.  The
mass differences between $\tilde{q}_L$ and $\tilde{b}$ are not large
enough for the peaks to separate and the result is a broad
distribution.

\subsection{Reconstruction of $\tilde\tau \tau$ Decays}

\begin{figure}[t]
\dofig{3.5in}{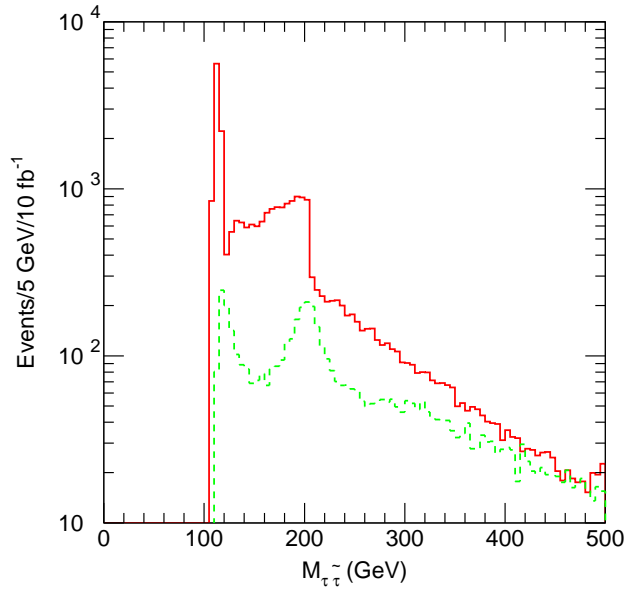}
\vskip-14pt
\caption{$\ttau$-$\tau$ mass distribution at Point G2b.
 Solid curve: Using visible
$\tau$ momentum. The dotted curve is obtained by selecting 
events where the missing $E_T$ is aligned with the tau direction by
$\Delta\phi< 0.1\pi$  and adding $\etmiss$ to $\tau$ momentum.
 \label{msleptau}}
\end{figure}

\begin{figure}[t]
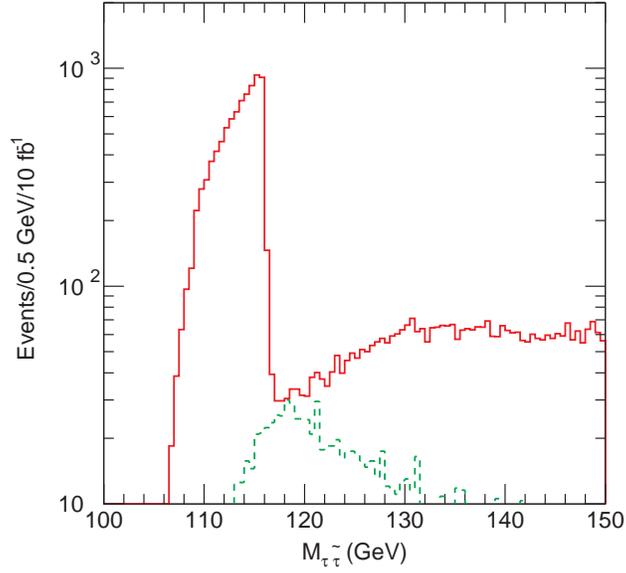

\dofig{3.5in}{g2b_msleptau1.ai}
\vskip-14pt
\caption{Same as Figure~\protect\ref{msleptau} on a finer scale with
  smaller bins.
\label{msleptau1}}
\end{figure}

The decay $\tchi_i^0 \to \ttau^\pm \tau^\mp$ is more difficult
to reconstruct than $\tchi_i^0 \to \tell_R^\pm \ell^\mp$, but it can
provide information on the gaugino content of the $\tchi_i^0$. 
A technique similar to that discussed in
Refs.~\citenum{point6} and \citenum{ianhtau}
can be used. Hadronic
$\tau$'s are selected with visible $p_T>20\,\GeV$ and $\eta<2.5$.
These are identified by taking jets where the number of charged tracks 
is less than or equal to three. The
simplest approach is to combine the visible $\tau$ momentum with the
slepton momentum. The resulting mass distributions are shown in
Figure~\ref{msleptau} and on an expanded scale for masses near the
$\lsp$ mass in Figure~\ref{msleptau1}\footnote{This plot 
requires at one least slepton, assumed 
  to be the ``muon candidate'' with the largest $p_t$, to be present in 
  the event. A second
  slepton may be required to facilitate triggering. Since all events
  have two sleptons, the efficiency for this is quite large. This
  additional requirement would reduce the rate shown in
  Figure~\ref{msleptau} slightly.}. These curves do not have true
peaks because of the missing $\nu_\tau$, but they do have fairly sharp
endpoints at the $\lsp$ and $\tchi_2^0$ masses. 

If the slepton momenta are included in the calculation of
$\etmiss$ and there are no other neutrinos, then $\etmiss$ can be used
to determine the true $\tau$ momentum. Only the highest $p_T$ $\tau$
is used, and the angle between it and the $\etmiss$ direction is
required to be $\Delta\phi < 0.1\pi$. The visible $\tau$ momentum is
then scaled by a factor $1 + \etmiss/E_{T,\tau}$, and the $\ttau$-$\tau$
mass is recomputed. This gives the dashed curves in
Figures~\ref{msleptau} and \ref{msleptau1}. As expected, including
$\etmiss$ not only reduces the statistics but also worsens the
resolution for $\lsp$, since the $\tau$ is very soft in this case.
However, it produces a peak near the right position for the
$\tchi_2^0$. While this peak probably does not improve the mass
resolution, it adds confidence that one is seeing a two-body
resonance. 

\subsection{Direct production of electro-weak sparticles}

\begin{figure}[t]
\dofig{3.5in}{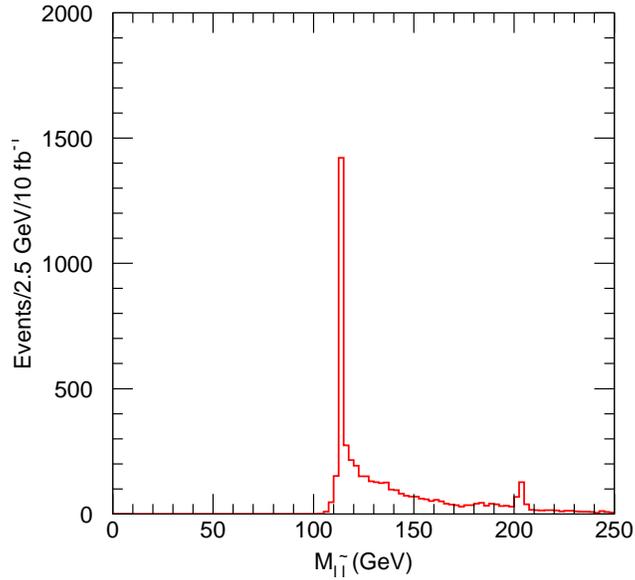}
\vskip-14pt
\caption{Same as Figure~\protect\ref{mchi} except that the additional
  requirement
$\Meff<100 $ GeV has been made.
\label{slep-lep}}
\end{figure}

We now return to the events present in the peak at low $\Meff$ shown 
in Figure~\ref{meff}.  As pointed out above these events are due to
the direct production of gauginos and sleptons. We begin with the
event sample used in Figure~\ref{mell} and remake that distribution
with the requirement that  $\Meff< 100 $ GeV. This is shown in
Figure~\ref{slep-lep}. This plot has a very strong peak at the mass of 
$\lsp$ and a weak, though still clear one, at
the mass of $\tchi_2^0$. This higher peak is suppressed as the lepton
from its decay is contributing to $\Meff$ and the cut throws away
some signal.

\begin{figure}[t]
\dofig{3.5in}{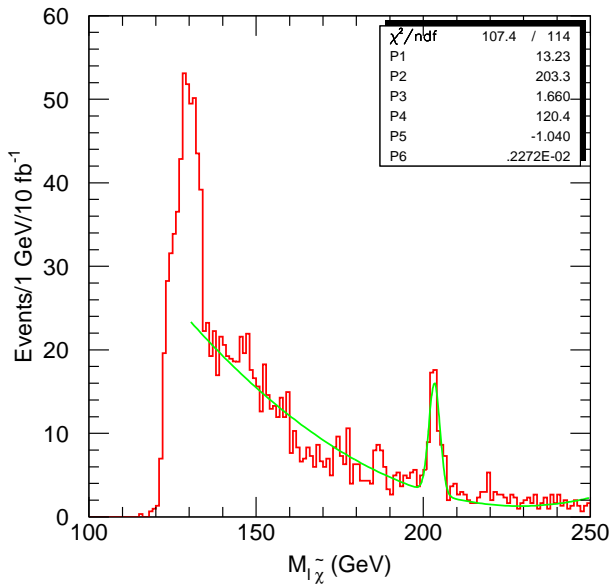 }
\vskip-14pt
\caption{The mass of the slepton-lepton-lepton system formed by
  selecting events in the mass range 110--120 GeV from 
Figure~\protect\ref{slep-lep} and combining this slepton-lepton pair
with and additional lepton. 
\label{slep-lep-lep}}
\end{figure}

We can also repeat the analysis of section~\ref{sec:slepton}, with the
addition of the cut $\Meff< 100 $ GeV.
We can take the events in the $\lsp$ peak (in the mass range 110--120
GeV)
of Figure~\ref{slep-lep} and 
then combine that reconstructed $\lsp$ momentum with an additional
charged lepton. The mass distribution of the resulting system is shown 
in Figure~\ref{slep-lep-lep}. This plot shows a peak at 204 GeV which
corresponds to the decay $\tilde{\ell}_L\to \lsp \ell \to
\tilde{\ell}_R^{\pm}\ell^\mp \ell$. The fit shown on this plot is a
linear combination of a exponential, a Gaussian and constant over the
mass range 135--250 GeV. The slepton peak is visible, although there
are many fewer events than in Figure~\ref{mchil3}. 
The kinematic feature from
the decay chain $\tilde{\chi}_1^\pm\to \tilde{\nu}\ell \to \lsp \nu 
\ell \to \tilde{\ell}_R^\pm\ell^\mp \nu\ell$
is still visible.

\section{Determining SUSY parameters}
\label{sec:scan}

Once a number of quantities have been measured, we can attempt to
determine the particular SUSY model and the values of the parameters.
The strategy will be to attempt to perform a global fit to the model
parameters using all of the available data, much as the Standard Model
is tested using the $W$ and $Z$ masses and the many quantities
precisely measured by LEP/SLC. Such a fit is beyond the scope of our
work, and we adopt a simpler procedure. We assume that from
measurements of global parameters such as those discussed in
section~\ref{sec:meff} we know the approximate scale of the superpartner
masses and have some idea that we might be in a GMSB model. The
object is then to determine the parameters of that model and check its
consistency. We must therefore determine the parameters $\Lambda$,
$M_m$, $N_5$, $\tan\beta$ and $\sgn\mu$ and $\Cgrav$.
If we know the value of one gaugino and one squark or slepton mass of
the first two generations
measured at the mass scale $M_m$ then $N_5$ and $\Lambda$ are
determined.
Since these are measured at the a lower energy scale, the physical
masses also depend on $M_m$ via the renormalization group scaling and we
need one more measurement to constrain it's value. Two gaugino masses 
and one  squark or slepton mass of the first two generations suffice.
$\tan\beta$ can be constrained either from the Higgs mass or from the
masses of the third generation squarks and sleptons. In the case of
model G2b, the splitting between $\tilde{e}_R$ and $\tilde{\tau}_1$
constrains it. Additional constraints on $\tan\beta$ and on $\sgn\mu$ 
arise from the Higgs, $\tchi_i^0$ and $\tchi_i^\pm$ masses.
 The only constraint upon $\Cgrav$
arises from the lifetime of the NLSP.
In cases G1a and G2a  and G2b we have precise measurements of the slepton and
$\chi_2^0$ and $\chi_1^0$ masses, so the less precise measurements of
the squark and gluino masses are not useful in determining the
fundamental parameters; they only provide powerful consistency checks.
Indeed the case G2b has so many observables, that it is enormously over 
constrained. 

In addition to the measurements presented above we assume that the
lightest Higgs boson has its mass determined precisely from its decay
to gamma-gamma. We will assume two values for its error; $\pm 3$~GeV,
which we estimate is the current systematic limit in the theoretical
calculations needed to relate it to the model parameters; and $\pm
100$~MeV which corresponds to the expected experimental precision.

Our strategy for determining the parameters is as follows. We choose a
point randomly in parameter space and compute the spectrum. We assign
a probability to this point determined from how well it agrees with
our ``measured quantities'' using our estimates of the errors on those
quantities. The process is repeated for many points and the
probabilities used to determine the central values of the parameters,
their errors and their correlations. $N_5$ is treated as a continuous
variable for these purposes.

At Point G1a the measurements discussed above for
10 (30) fb$^{-1}$ where statistical errors will still be important, {\it viz.}
\begin{itemize}
\item 
$M_{\tchi_2^0} \sqrt{1-\left(M_{\tell_R} \over M_{\tchi_2^0}\right)^2}
\sqrt{1-\left(M_{\lsp} \over M_{\tell_R}\right)^2} = 105.1 \pm 0.10 (0.10)\,\GeV$ ,
 \item $\sqrt{M_{\tchi_2^0}^2 -
M_{\chi_1^0}^2}= 189.7\pm 0.50 (0.30)$ GeV,
\item $\sqrt{M_{\ell_R}^2-M_{\chi_1^0}^2}= 112.7 \pm .20 (0.15)$ GeV
\item $\sqrt{M_{\chi_2^0}^2-M_{\ell_R}^2}= 152.6\pm .50 (0.30)$ GeV
\item $m_{h^0}= 109.47\pm 3$ GeV,
\end{itemize}
imply that
\begin{itemize}
\item 
$\Lambda= 90000 \pm 1800 (1200)$ GeV, 
\item $M_m= 500000 \pm 150000 (120000)$ GeV,
\item $\tan \beta= 5.0 \pm 1.5 (0.9)$,
\item $N_5 =  1 \pm 0.012 (0.10)$. 
\end{itemize}
$\sgn\mu$ is determined unambiguously. Some improvement will be
possible with greater integrated luminosity until the systematic limit 
is reached. This will occur for
100 fb$^{-1}$ of integrated luminosity.
 Assuming that the errors are then
$\pm 100$ MeV, $\pm 200$ MeV, $\pm 100$ MeV, $\pm 200$ MeV and $\pm 3$
GeV respectively, the uncertainties on the parameters reduce:
\begin{itemize}
\item $\Lambda= 90000 \pm 630$ GeV, 
\item $M_m= 500000 \pm 80000$ GeV,
\item $\tan \beta= 5.0 \pm 0.3$,
\item $N_5  =  1 \pm 0.008$.
\end{itemize}
If
the error on the Higgs mass is reduced to $\pm 100$~MeV, the uncertainty  on
$\tan\beta$
reduces to $\pm 0.1$; the other uncertainties are unchanged.
The poorer precision on $M_m$ reflects the fact 
that it enters only via the renormalization group evolution and
therefore that the observed masses depend only logarithmically upon
it. 

At Point G1b, we have for 10 fb$^{-1}$ of integrated luminosity
\begin{itemize}
\item 
$M_{\tchi_2^0} \sqrt{1-\left(M_{\tell_R} \over M_{\tchi_2^0}\right)^2}
\sqrt{1-\left(M_{\lsp} \over M_{\tell_R}\right)^2} = 105.1 \pm 0.10 \,\GeV$,
\item $m_{\tilde{g}}-m_{\tchi_2^0}= 523 \pm 50$ GeV,
\item $m_{h^0}= 109.47 \pm 3$ GeV,
\end{itemize}
The precision on the second of these numbers can be expected to
increase with more integrated luminosity; the others are systematics limited.
These are sufficient only to constrain the following with any degree
of precision
\begin{itemize}
\item 
$\Lambda N_5= 90000 \pm 860 $ GeV, 
\item $\tan \beta= 5.0^{+1.9}_{-1.3}$ 
\end{itemize}
This is due to  an accident in our choice of parameters. The
position of the kinematic endpoint of Figure~\ref{g1bdilep} is
insensitive to variations of the slepton mass, 
when $m_{\tilde{\ell}_R}=\sqrt{m_{\tchi_2^0}m_{\lsp}}$. For our choice of 
parameters these quantities differ by 0.5 GeV! As $\lsp$ and
$\tchi_0^2$ are almost purely gaugino, these relations provide
constraints 
only on the gaugino masses, {\it i.e.} on the product  $\Lambda N_5$.
If we assume that the we are able to constrain the average light
squark mass within 50 GeV of its nominal value as appears to be
possible from the discussion surrounding  
Figure~\ref{g1blj}, for 30 fb$^{-1}$, we then obtain
\begin{itemize}
\item $\Lambda N_5= 90000 \pm 620$ GeV,
\item $\Lambda = 90000 \pm 8100$ GeV,
\item $M_m < 7\times 10^8$  GeV (95\% confidence)
\item $\tan \beta= 5.0^{+1.9}_{-1.3}$ 
\end{itemize}
Here we have restricted $M_m> \Lambda$; the model is not sensible if
this is not the case. If
the error on the Higgs mass is reduced to $\pm 100$ MeV, the uncertainty  on
$\tan\beta$
reduces to $\pm 0.1$; the other uncertainties do not reduce. 
It is not possible to determine
$\sgn\mu$ using the signals that we have shown. For example, the set
of parameters $\Lambda=104500$ GeV, $M_m=0.239\times 10^9$ GeV,
$N_5=0.872$ and $\sgn\mu=-1$ is acceptable. The mass of $\tchi_2^0$
($\lsp$) 
is increased by 19 (6) GeV and the mass of $\tilde{e}_R$ by 40 GeV, but 
this case has the same end point in Figure~\ref{g1bdilep}. An
independent constraint on the slepton mass or a measurement of the
squark mass with a precision of order 10 GeV is needed to eliminate
this case.

At Point G2a the measurements discussed above for 10 fb$^{-1}$ {\it viz.}
\begin{itemize}
\item $\sqrt{m_{\tilde{\chi}_1^0}^2-m_{\tilde{e}_R}^2}= 52.21\pm 0.07$ GeV,
\item $\sqrt{m_{\tilde{\chi}_2^0}^2-m_{\tilde{\mu}_R}^2}= 175.94 \pm 0.27$ GeV,
\item $\sqrt{m_{\tilde{q}_R}^2-m_{\tilde{e}_R}^2 }= 640 \pm 7$ GeV,
\item $\sqrt{m_{\tilde{q}_R}^2-m_{\tilde{e}_R}^2 }
\sqrt{1- {M_{\tell_R}^2\over M_{\lsp}^2}}/\sqrt{M_{\tq_R}^2 - M_{\lsp}^2} 
=0.45 \pm 0.004$ 
 \item $m_{h^0}= 106.61 \pm 3$ GeV,
\end{itemize}
imply that
\begin{itemize}
\item $\Lambda= 30000 \pm 410$ GeV,  
\item $M_m= 250000 \pm 44000$ GeV,
 \item $\tan \beta= 5.0 \pm 0.7$
\item $N_5       =  3 \pm 0.036$. 
\end{itemize}
$\sgn\mu$ is determined unambiguously.  Only small improvements can be 
expected as the integrated luminosity is increased above 10
fb$^{-1}$. 
If we reduce the errors on the first two quantities to 50 and 180 MeV
respectively, their likely systematic limits, as might be achieved with 30 fb$^{-1}$ of data we
obtain
\begin{itemize}
\item $\Lambda= 30000 \pm 380$ GeV,  
\item $M_m= 250000 \pm 42000$ GeV,
\item $\tan \beta= 5.0 \pm 0.7$
\item $N_5       =  3 \pm 0.035$. 
\end{itemize}
If the error on the Higgs mass were reduced to $\pm 100$ MeV, the error on
$\tan\beta$
reduces to $\pm 0.04$. 

At Point G2b the measurements discussed above {\it viz.}
\begin{itemize}
\item $m_{\tilde{e}_R}= 102.67 \pm 0.1$ GeV,
\item $m_{\tilde{\mu}_R}= 102.67 \pm 0.1$ GeV,
\item $m_{\tilde{\tau}_1}= 101.35 \pm 0.1$ GeV,
\item $m_{\lsp}= 115.18 \pm 0.1$ GeV,
\item $m_{\tilde{\chi}_2^0}= 203.71 \pm 0.2$ GeV, and
 \item $m_{h^0}= 106.61 \pm 3$ GeV,
\end{itemize}
imply that
\begin{itemize}
\item $\Lambda= 30000 \pm 175$ GeV  
\item $M_m= 250000 \pm 22500$
 \item $\tan \beta= 5.0 \pm 0.21$
\item $N_5       =  3 \pm 0.014$. 
\end{itemize}
$\sgn\mu$ is determined unambiguously. These measurements are likely
to be at their systematic limits with 10 fb$^{-1}$ of integrated
luminosity and further improvement will be difficult.  If
the error on the Higgs mass is reduced to $\pm 100$ MeV, the error on
$\tan\beta$
reduces to $\pm 0.02$. 

In case G1b whose global signatures are similar to SUGRA, we must
address the issue of whether it could be confused with a SUGRA model.
We search SUGRA parameter space for test of parameters that could be 
consistent with the following:
\begin{itemize}
\item 
$M_{\tchi_2^0} \sqrt{1-\left(M_{\tell_R} \over M_{\tchi_2^0}\right)^2}
\sqrt{1-\left(M_{\lsp} \over M_{\tell_R}\right)^2} = 105.1 \pm 0.10 \,\GeV$,
\item $m_{\tilde{g}}-m_{\tchi_2^0}= 523 \pm 50$ GeV,
\item $m_{h^0}= 109.47 \pm 3$ GeV,
\item $m_{\tilde{q}}-m_{\tilde{g}}=200 \pm 500$ GeV.
\end{itemize}
We obtain the following solution for the SUGRA parameters
\begin{itemize}
\item 
$m_0=100 \pm 20$ GeV 
\item $m_{1/2}=295 \pm 6$ GeV
\item $\tan \beta=4.5 \pm 1.1$
\item $\sgn\mu=+1$
\item $A=250\pm 200$ GeV
\end{itemize}
This solution has only a 15\% probability. The central value of the light
squark masses for this solution is 760 GeV. Most of the other masses
are similar to those of case G1b with the exception of $\tilde{e}_L$
which has mass 100 GeV larger. This result illustrates the general
difference between SUGRA and GMSB models. The mass splitting between 
the squarks and 
$\tilde{e}_R$ is larger in the GMSB case. If we assume that
$m_{\tilde{q}}-m_{\tilde{g}}=200 \pm 75$ GeV  as we might expect from
the methods of Section~\ref{sec:g1b} with at least 30 fb$^{-1}$ of
integrated luminosity, then the SUGRA solution is
eliminated (it has $10^{-5}$ probability) and the ambiguity resolved.

\section{Discussion and Conclusions}

In this paper we have given examples of how LHC experiments might
analyze supersymmetry events if SUSY exists and if
the pattern of superpartner masses is given by gauge mediated models
of supersymmetry breaking.
We have illustrated the four classes of phenomenology to be expected in
such models: events with missing energy that are similar to those
expected in SUGRA models
 (Point G1b); events with a pair of isolated photons from
$\lsp$ decay (Point G1a); events with long lived sleptons (Point G2b);
 and events with leptons 
from prompt slepton decay (Point G2a). In the first case, we have discussed how
measurements can be made which enable one to prove that  the gauge
mediated model and not SUGRA is responsible for the pattern of masses.
In the other cases, detection and measurement is easier than in SUGRA.
Characteristic features are present such as photons, leptons or
stable charged particles, that enable the supersymmetry signals to be
extracted trivially from Standard Model backgrounds. In addition,
these features make it possible to identify and use longer decay
chains.

We have illustrated a technique whereby the supersymmetry events can be 
fully reconstructed despite the presence of two undetected particles.
The technique relies upon there being a decay chain of sufficient
length that occurs twice in the same event. Each step in the chain
provides a constraint and sufficient constraints can occur that
together with the measurement of $\etmiss$ enables the event to be
reconstructed. In such cases the masses of the superparticles in the decay
chain can be measured directly rather than inferred from kinematic 
distributions.

In all of the cases discussed in this paper, as in the SUGRA cases
discussed previously \cite{hinch,previous,fabiola}, the LHC will be
capable of making many precise measurements that will enable the
underlying model of supersymmetry to be severely constrained should
supersymmetric particles be observed. The key to all of these
analyses is the ability to identify a characteristic final state
arising from the decay of a sparticle that is copiously produced. The
main such decays that have been used in the SUGRA and GMSB analyses
done so far are:
\begin{itemize}
\item   Dileptons from $\tchi_2\to \lsp \ell^+\ell^-$ or $\tchi_2\to
\tilde{\ell_R} \ell_R \lsp \ell^+\ell^-$; 
\item   taus from the decays $\tchi_2\to \lsp
\tilde{\tau_1}^+\tau_1^-$ or $\tchi_2\to \tilde{\tau_1} \tau_1 \to
\lsp \tau^+\tau^-$, which can dominate when $\tan\beta$ is large;
\item   $\tchi_2\to\lsp h \to \lsp b\overline{b}$.
\end{itemize}
In the case of GMSB or $R$-parity breaking models the subsequent decay
of $\lsp$ can provide additional information and constraints. Of
course the information extracted in this way is only a small fraction
of the total available. A complete analysis will involve generating
large samples of events for many SUSY models and comparing all
possible distributions with experiment. 

\section*{Acknowledgements}

We are particularly grateful to John Hauptman for a discussion about
the usefulness of multiple mass constraints during the early stages of
this work.

This work was supported in part by the Director, Office of Energy
Research, Office of High Energy Physics, Division of High Energy
Physics of the U.S. Department of Energy under Contracts
DE-AC03-76SF00098 and DE-AC02-98CH10886.  Accordingly, the U.S.
Government retains a nonexclusive, royalty-free license to publish or
reproduce the published form of this contribution, or allow others to
do so, for U.S. Government purposes. 

\newpage
\section*{Appendix}
The details of the full reconstruction used in section \ref{sec:recon}
are given here.

        Events are selected to have four leptons and two photons with
a unique combination of two leptons and one photon coming from each
$\tchi_2^0$ decay. The $\tchi_2^0$, $\tell_R$, and $\lsp$ masses are
assumed to be known precisely from the distributions discussed in
section~\ref{sec:g1alep},
 and the $\tG$ mass was assumed to be (essentially)
zero. This leads to the following set of equations for the 4-momentum
$p$ of the gravitino:
\begin{eqnarray*}
p^2 &=& 0\\
(p + p_\gamma)^2 &=& M_{\lsp}^2\,,\\
(p + p_\gamma + p_{\ell_2})^2 &=& M_{\tell}^2\,,\\
(p + p_\gamma + p_{\ell_2} + p_{\ell_1})^2 &=& M_{\tchi_2^0}^2\,.
\end{eqnarray*}
This implies that the gravitino momentum $p$ is given in terms of the
photon momentum $k$ and the lepton momenta $l$ and $q$ by
\begin{eqnarray*}
2p_0k_0 - 2\vec p \cdot \vec k &=& M_{\lsp}^2 \equiv C_k\,, \\
2p_0l_0 - 2\vec p \cdot \vec l &=& M_{\tell_R}^2 -M_{\lsp}^2
-2k \cdot l \equiv C_l\,, \\
2p_0q_0 - 2\vec p \cdot \vec q &=& M_{\tchi_2^0}^2 -M_{\tell_R}^2 
-2(k+l)\cdot q -q^2 \equiv C_q\,,
\end{eqnarray*}
where
$$
p_0 = \sqrt{\vec p}\,.
$$
It is now clear  that they
give two linear and one quadratic constraint and hence an additional
$2\times2$ ambiguity. The solution is straightforward. From the above
equations one finds
\begin{eqnarray*}
\vec p \cdot \vec{D}_1 &=& E_1\,, \\
\vec p \cdot \vec{D}_2 &=& E_2\,,
\end{eqnarray*}
where
\begin{eqnarray*}
\vec{D}_1       &=& 2 l_0 \vec{k} - 2 k_0 \vec{l}\,, \\
E_1             &=& -l_0 C_k + k_0 C_l\,, \\
\vec{D}_2       &=& 2 q_0 \vec{k} - 2 k_0 \vec{q}\,, \\
E_1             &=& -q_0 C_k + k_0 C_q\,.
\end{eqnarray*}
These can be solved to give
$$
p_i = F_i + G_i p_z, \quad i=1,2\,,
$$
where
\begin{eqnarray*}
F_x &=& {E_1D_{2y}-E_2D_{1y} \over D_{1x}D_{2y}-D_{2x}D_{1y}}\,, \\
G_x &=& -{D_{1z}D_{2y}-D_{2z}D_{1y} \over D_{1x}D_{2y}-D_{2x}D_{1y}}\,, \\
F_y &=& {E_1D_{2x}-E_2D_{1x} \over D_{1y}D_{2x}-D_{2y}D_{1x}}\,, \\
G_y &=& -{D_{1z}D_{2x}-D_{2z}D_{1x} \over D_{1y}D_{2x}-D_{2y}D_{1x}}\,.
\end{eqnarray*}
This is then substituted into the first of the original equations to
yield a quadratic equation for $p_z$, 
$$
H_0 + H_1 p_z + H_2 p_z^2 = 0\,,
$$
where
\begin{eqnarray*}
H_0 &=& 4k_0^2(F_x^2+F_y^2+M_\lsp^2)-(C_k+2F_xk_x+2F_yk_y)^2\,, \\
H_1 &=& 4k_0^2(2F_xG_x+2F_yG_y)-2(C_k+2F_xk_x+2F_yk_y)
(2G_xk_x+2G_yk_y+2k_z)\,,\\
H_2 &=& 4k_0^2(G_x^2+G_y^2+1)-(2G_xk_x+2G_yk_y+2k_z)^2\,.
\end{eqnarray*}
This gives two solutions for the $\tG$ momentum $p$ for a given
assignment of the other momenta provided $H_1^2-4H_0H^2 \ge 0$ and no
solution otherwise.


\end{document}